\documentclass[lettersize,journal]{IEEEtran}
\usepackage{amsmath,amsfonts}
\usepackage{algorithmic,amsfonts,amsmath,amssymb,booktabs,cite,color,epsfig,float,graphicx,latexsym,makecell,multirow,palatino,soul,threeparttable,bm,subfig}
\usepackage{mathrsfs,amsthm}
\usepackage{lineno}
\usepackage[ruled,linesnumbered]{algorithm2e}
\newtheorem{defi}{\textbf{Definition}}
\newtheorem{theo}{\textbf{Theorem}}

\SetKwRepeat{Do}{do}{while}

\begin{document}

\title{A Privacy-Preserving Data Collection Method for Diversified Statistical Analysis}

\author{Hao Jiang,
	    Quan Zhou,
	    Dongdong Zhao,
	    Shangshang Yang, 
	    Wenjian Luo,~\IEEEmembership{Senior Member,~IEEE},
	    and Xingyi Zhang,~\IEEEmembership{Senior Member,~IEEE}\\
        % <-this % stops a space
	\thanks{This study is partly supported by the National Natural Science Foundation of China (No. 62372001). (\textit{Corresponding author: Wenjian Luo and Xingyi Zhang})}
	\thanks{H. Jiang and X. Zhang are with the Key Laboratory of Intelligent Computing and Signal Processing of the Ministry of Education, School of Computer Science and Technology, Anhui University, Hefei 230601, China (email: haojiang@ahu.edu.cn; xyzhanghust@gmail.com).}
	\thanks{Q. Zhou and S. Yang are with the Key Laboratory of Intelligent Computing and Signal Processing of the Ministry of Education, School of Artificial Intelligence, Anhui University, Hefei 230601, China (email: 3467550194@qq.com, yangshang0308@gmail.com).}
	\thanks{D. Zhao is with the School of Computer Science and Artificial Intelligence, Wuhan University of Technology, Wuhan, Hubei, China (email: zdd@whut.edu.cn).}
	\thanks{W. Luo is with the Guangdong Provincial Key Laboratory of Novel Security Intelligence Technologies, School of Computer Science and Technology, Harbin Institute of Technology, Shenzhen 518055, Guangdong, China (email: luowenjian@hit.edu.cn).}

}

\markboth{IEEE Transactions on XXX,~Vol.~xxx, No.~xxx}%
{Shell \MakeLowercase{\textit{et al.}}:}


\maketitle

\begin{abstract}
%Privacy-preserving methods based on data perturbation have been widely adopted in various scenarios for sensitive data collection due to their efficiency and the elimination of the need for a trusted third party. 
%However, existing perturbation-based methods primarily focus on ensuring data usability through individual statistical indicators, neglecting the quality of collected data from the perspective of data distribution. 
%Consequently, these methods often fail to meet the diverse statistical analysis requirements commonly encountered in practical data analysis. 
%As a promising sensitive data perturbation method, negative survey is able to complete the task of collecting and distributing sensitive information while protecting users' personal privacy.
%However, existing methods are only designed for discrete sensitive information and are difficult to apply to the task of collecting real valued data distributions.
%To address this issue, for the first time, this paper proposes a real-value negative survey model, called RVNS, for real-value sensitive information collection. 
%Compared to existing perturbation-based methods, the RVNS not only effectively protects the privacy of users' real-valued data but also ensures data usability from the perspective of the overall data distribution. 
%Experimental results on both synthetic and real-world datasets demonstrate that the proposed method outperforms four state-of-the-art perturbation-based privacy-preserving methods in terms of data utility at the same level of privacy preservation.
Data perturbation-based privacy-preserving methods have been widely adopted in various scenarios due to their efficiency and the elimination of the need for a trusted third party. 
However, these methods primarily focus on individual statistical indicators, neglecting the overall quality of the collected data from a distributional perspective. 
Consequently, they often fall short of meeting the diverse statistical analysis requirements encountered in practical data analysis. 
As a promising sensitive data perturbation method, negative survey methods is able to complete the task of collecting sensitive information distribution while protecting personal privacy. 
Yet, existing negative survey methods are primarily designed for discrete sensitive information and are inadequate for real-valued data distributions.
To bridge this gap, this paper proposes a novel real-value negative survey model, termed RVNS, for the first time in the field of real-value sensitive information collection. 
The RVNS model exempts users from the necessity of discretizing their data and only requires them to sample a set of data from a range that deviates from their actual sensitive details, thereby preserving the privacy of their genuine information. 
Moreover, to accurately capture the distribution of sensitive information, an optimization problem is formulated, and a novel approach is employed to solve it. 
Rigorous theoretical analysis demonstrates that the RVNS model conforms to the differential privacy model, ensuring robust privacy preservation. 
Comprehensive experiments conducted on both synthetic and real-world datasets further validate the efficacy of the proposed method. 
The results indicate that, when compared to four state-of-the-art perturbation-based privacy-preserving methods specifically designed for real-value data, the RVNS achieves superior data utility for diversified statistical analysis while maintaining the same level of privacy preservation. 

\end{abstract}


\section{Introduction}
\IEEEPARstart{I}{n} the digital era, the ease of collecting and preserving information has laid a solid foundation for data mining methodologies \cite{4490259}. 
However, this convenience is accompanied by an increasingly pressing concern regarding data privacy. 
The frequent occurrence of privacy breaches has led some users to distrust the ability of data collectors to safeguard their personal data privacy, resulting in a reluctance to provide sensitive personal information \cite{7823333}. 
To alleviate the concerns of users about privacy breaches and better protect their data, various sensitive data collection methods that do not require a trusted third party (TTP) have been proposed \cite{7163223}. 
These include methods based on encryption \cite{10261340}, anonymization \cite{10409614}, and perturbation \cite{9658125perturbation}, among others. 
Among these approaches, perturbation-based methods have garnered extensive attention due to their wide range of applications and high computational efficiency \cite{chamikara2018efficient}.

Despite their promising performance, most existing perturbation-based methods do not focus on similarity in data distribution before and after perturbation, so they usually can only ensure the accuracy of only particular statistical indicators, such as $Mean$ \cite{barbosa2014lightweight}. 
In practical applications, the demand for statistical analysis is often diversified. 
For instance, analyzing the physical condition of people in a region may require statistical indicators like average height, weight, body mass index (BMI), and prevalence of certain health conditions \cite{petersen2019health}. 
Conversely, examining income levels in the same area might necessitate statistics on median income, income distribution, Gini coefficient, and percentage of individuals falling below the poverty line \cite{tai2023user}.
%Therefore, there is an urgent need for a sensitive data collection method that can flexibly support a wide range of statistical analyses.%
Therefore, a sensitive data collection method that can flexibly support a wide range of statistical analyses is urgently needed.

The ability to perform diverse statistical analyses hinges on accurately obtaining the overall distribution of all users' sensitive information. 
By sampling from this distribution, relevant data analysis tasks can be accomplished. 
Hence, if we can acquire the accurate overall distribution of data while protecting personal sensitive information privacy, we can establish a data collection method tailored for diversified statistical analyses.
However, the data collected by existing perturbation-based methods cannot guarantee accurate distribution estimation \cite{turgay2023perturbation}.
Negative surveys emerge as a promising sensitive information collection method capable of gathering the distribution of sensitive information while preserving individual privacy \cite{jiang2019privacy}. 
However, most existing negative survey methods are designed for discrete data and are challenging to apply directly to continuous data. 
Although some approaches enable negative surveys to handle continuous sensitive information by discretizing the data \cite{jiang2019privacy}, this discretization process leads to a loss of original data precision, thereby affecting data usability, particularly in terms of statistical characteristics such as variance.

In this paper, we propose a continuous data collection method for diversified statistical analysis based on the negative survey model. 
In this method, individual users only need to return a set of sampled data different from their actual sensitive information. 
The data collector can then utilize specific methodologies to reconstruct the distribution of all users' original information from the collected samples. 
Throughout this process, each user does not disclose their true information, while the collector obtains the distribution of all users' sensitive information. 
Consequently, our proposed method facilitates the collection of sensitive information while protecting personal privacy and supports diversified statistical analyses.

The specific contributions of this paper are summarized as follows.
\begin{enumerate}
	\item A real-value negative survey model, termed RVNS, is first proposed to safeguard the privacy of continuous sensitive information. In this model, users are exempted from the necessity of discretizing their data. Instead, they are required solely to sample a set of data from the range of sensitive information, which deviates from their actual sensitive details. As users do not disclose genuine sensitive information, the privacy of this information is preserved.
	\item To accurately obtain the distribution of sensitive information, an optimization problem is formulated to model the reconstruction process. Subsequently, a novel optimization approach is utilized to address this problem. The precise distribution obtained through this method enables the collected data to support a wide range of statistical analyses.
	\item Rigorous theoretical analysis and comprehensive experiments are conducted to ascertain the efficacy of the RVNS. The theoretical analysis demonstrates that the RVNS conforms to the differential privacy model. Furthermore, the experimental results indicate that, when compared to four perturbation-based methods specifically designed for continuous data, the RVNS achieves superior utility while maintaining the same level of privacy preservation.	
\end{enumerate}

The remainder of this paper is structured as follows. Section \ref{sec:relate} provides a comprehensive review of existing perturbation-based methods that do not necessitate a TTP, introduces the concept of negative surveys concisely, and gives the motivation of this paper. Section \ref{sec:method} offers detailed insights into the developed RVNS model and Section \ref{sec:analy} analyzes the developed RVNS. In Section \ref{sec:exper}, the performance of the proposed RVNS is rigorously validated through experimental evaluations. Lastly, Section \ref{sec:conclu} presents the concluding remarks of this paper.

\section{Related work}
\label{sec:relate}
\subsection{Existing Perturbation-based Methods for Real-value Data Without a TTP}
Existing perturbation-based methods for real-value data without a TTP can be broadly classified into two main groups according to the perturbation techniques utilized.
The first category of methods involves perturbing data by introducing noise into the original data. 
These methods aim to obscure sensitive data through noise addition while preserving certain data properties, thus maintaining specific statistical properties and protecting data privacy.
For example, Zheng et al. \cite{zheng2022decentralized} proposed a distributed noise addition method designed for electricity consumption data. 
This method injects Laplacian noise into the data of each client in a distributed manner and randomly shuffles the processed data to ensure that the aggregated data still satisfies differential privacy. 
To address the issue of unbounded noise injection in existing methods like the Laplacian mechanism, Wang et al. \cite{wang2019collecting} designed a Piecewise Mechanism for noise addition. 
This method uses a randomized perturbation function to add noise to the original data, ensuring that the perturbed data remains within the value range of the original data.
Although noise addition techniques effectively conceal individual-level information, making it difficult for adversaries to infer private details from the data, the introduction of noise increases data dispersion, potentially affecting the accuracy of estimating distributions from the collected data.

The second category utilizes randomization techniques to perturb data. 
These methods distort the original data by incorporating random elements, thereby protecting data privacy. 
Subsequently, collectors employ specific statistical methods (hereinafter referred to as reconstruction methods) to analyze the distorted data and obtain the distribution of sensitive information.
For example, Sei et al. \cite{sei2017differential} devised a sensitive data collection method in the context of mobile crowdsensing. 
This method first discretizes continuous data into multiple bins and then employs random response technology to probabilistically perturb users' true data into other bins. 
Afterward, specific reconstruction methods are used to estimate the frequency of users belonging to different bins. 
To mitigate the impact of discretization on data usability, Kulkarni et al. \cite{kulkarni2019answering} introduced a sensitive data collection method based on hierarchical histograms. 
This method groups users and discretizes their data at different granularities, employing random response technology at each granularity level to perturb the data. 
Collectors estimate the frequency distribution of the data by leveraging the relationships between different granularities, thereby enhancing data usability.
Although randomization techniques allow for the collection of distributions of sensitive information, they require prior discretization of the data, with bins that cannot be altered once determined. 
Furthermore, such methods overlook the differences among data within the same bin, limiting the usability of the obtained data distributions. 
Thus, while both categories of methods significantly contribute to data privacy preservation, they also present specific challenges and limitations that require further research and innovation.

\subsection{Negative Survey}
Negative surveys have emerged as a promising method to collect data, offering a nuanced approach to gathering sensitive data while safeguarding individuals' privacy \cite{esponda2009surveys,esponda2006negative}. 
Specifically, negative surveys operate on the principle of eliciting information by requiring individuals to provide a category that does not correspond to their true situation, referred to as the 'negative category.' 
For instance, in collecting salary information, a typically direct question might be posed:

\textbf{Q1}: \textit{What is your monthly salary range?}

\textit{\quad \quad A. Below 3000 yuan }
	
	\textit{\quad \quad B. 3000-5000 yuan }
	
	\textit{\quad \quad C. 5000-7000 yuan }
	
	\textit{\quad \quad D. 7000-8000 yuan }
	
	\textit{\quad \quad E. Above 8000 yuan}

Responding to such a question directly reveals one's salary range, thereby compromising privacy. 
Conversely, in a negative survey, the question is framed as:

\textbf{Q2}: \textit{Which of the following salary ranges does your monthly salary not fall into?}

\textit{\quad \quad A. Below 3000 yuan }

\textit{\quad \quad B. 3000-5000 yuan }

\textit{\quad \quad C. 5000-7000 yuan }

\textit{\quad \quad D. 7000-8000 yuan }

\textit{\quad \quad E. Above 8000 yuan}

A respondent earning 6000 yuan per month would randomly select from options A, B, D, or E, thereby only denying their true category. 
This mechanism ensures that the respondent's actual salary remains undisclosed, thus preserving privacy.
Collectors subsequently utilize sophisticated statistical techniques, known as reconstruction methods, to derive the frequency distribution of the original information from the negative responses. 
These methods allow for the indirect inference of sensitive data patterns without compromising individual privacy.
The versatility of negative surveys has led to their adoption across various domains for collecting sensitive data, including location information \cite{jiang2017novel}, health data \cite{jiang2019consistency}, and rating information \cite{luo2017rating}. 
This versatility underscores the method's adaptability and potential for widespread application in fields where privacy concerns are paramount.
However, negative surveys are primarily designed for categorical sensitive data. 
While real-valued data can be processed through discretization \cite{jiang2019privacy}, this often results in a loss of data granularity and, consequently, limited data usability. 

\subsection{Motivation}
From the above analysis, it can be found that existing perturbation-based methods have demonstrated their efficacy in preserving individual privacy and capturing the overall distribution of sensitive information across a user population. 
However, these approaches often suffer from limited data distribution precision, which hampers the subsequent analytical requirements, especially when the specific data analysis methodologies are yet to be determined.
Recently, Li et al. \cite{li2020estimating} introduced a method for collecting real-valued sensitive data based on a randomized response technique. 
This method circumvents the need for prior discretization of data and enables the acquisition of the overall distribution of real-valued data. 
Nonetheless, it is accompanied by three notable drawbacks. 
Firstly, users still have a considerable probability of disclosing their original sensitive information, leading to potential privacy concerns. 
This issue is particularly acute when collecting highly sensitive information, such as illegal activities, where users may be skeptical of the claimed confidentiality measures by the collectors. 
Secondly, while the aggregator can posteriorly partition data into various bins as required, the inherent need for data discretization persists. 
This necessitates overlooking the intrinsic differences among data within the same bin, thereby constraining the utility of the obtained data distributions.
Thirdly, the proposed reconstruction method independently focuses on the probability density of different sampling points, neglecting the holistic data distribution. 
Consequently, the resulting data distribution lacks precision.
In contrast, negative surveys, as a sensitive information collection technique, offer a promising alternative by enabling the acquisition of the true distribution of sensitive information without requiring users to disclose their actual data. 
This characteristic makes negative surveys more palatable to users. 
However, the existing negative survey models are not readily adaptable for collecting the distribution of real-valued data.

Given these limitations, this paper presents a real-value data collection method tailored for diverse statistical analyses based on the negative survey model. 
Our approach models the reconstruction process from the perspective of the overall data distribution, thereby enhancing the accuracy of the collected data distribution. 
By leveraging the principles of negative surveys and incorporating advanced statistical techniques, we aim to strike a balance between privacy preservation and data utility, addressing the precision shortcomings of existing methods. 
The details of our approach will be given in the next section.

\section{The Proposed RVNS}
\label{sec:method}
\subsection{Problem Statement}
\label{sec:state}

\begin{figure}[t]
    \centering
    \includegraphics[width=0.5\textwidth]{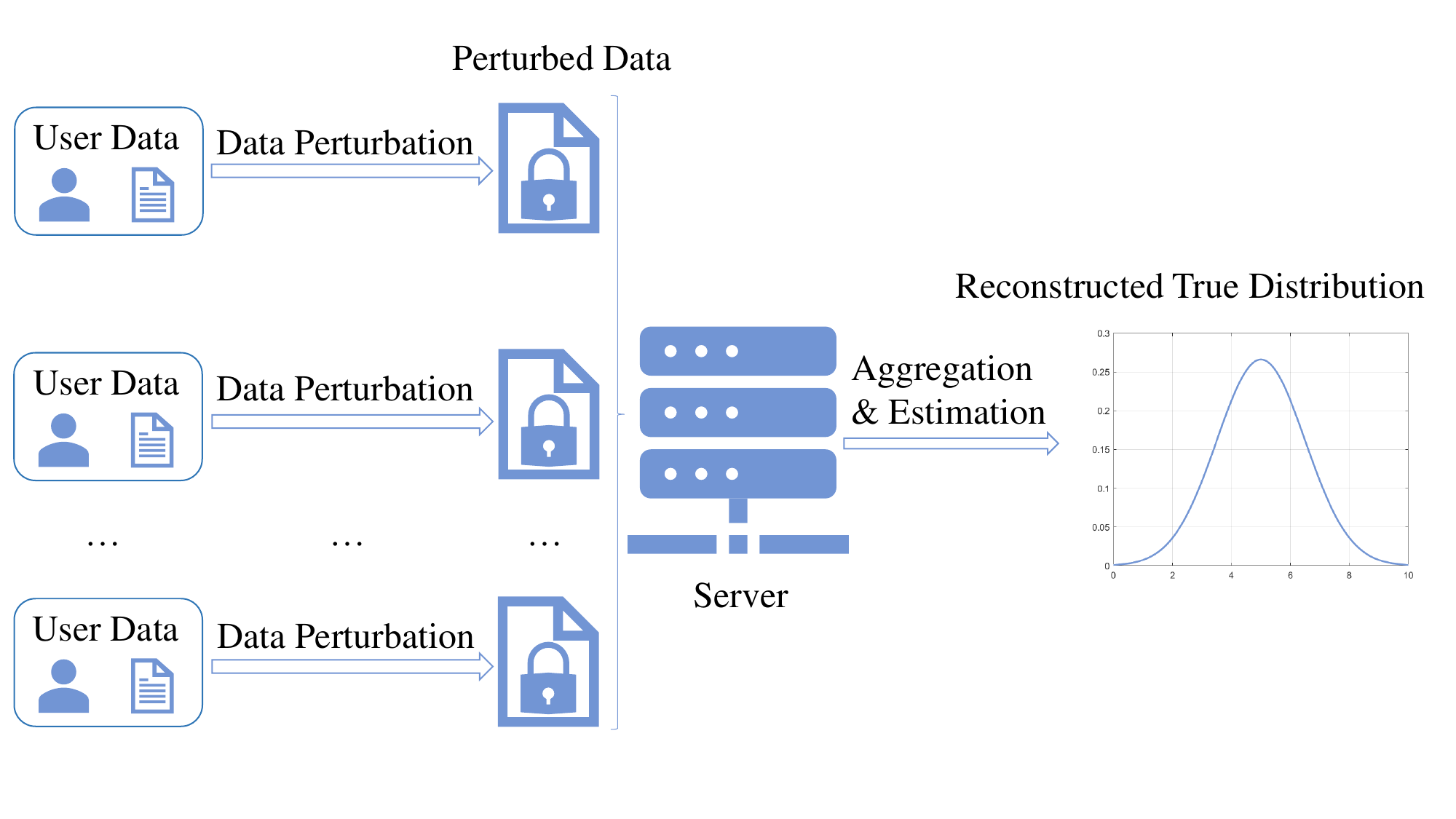}
    \caption{System model.}
    \label{fig:framework}
\end{figure}

In this paper, we delve into a client-server architecture that resembles numerous local differential privacy based frameworks, consisting of a server and an ensemble of participating users. 
The primary objective of this architecture is to ensure privacy preservation. 
To achieve this, users locally obfuscate their real-valued data on their client devices according to a predefined perturbation protocol. 
Subsequently, these perturbed data are transmitted to the server without any additional interaction. 
The server, functioning as the service provider for a specific application, aggregates the received perturbed data from users and estimates the underlying data distribution. 
A graphical illustration of the general system model is presented in Figure \ref{fig:framework}.

\begin{table}[t]
\centering
\caption{List of notations.}
\label{tab:symbols}
\begin{tabular}{>{\centering\arraybackslash}m{0.15\columnwidth}|>{\raggedright\arraybackslash}p{0.7\columnwidth}}
\hline
\textbf{Notation} & \textbf{Description} \\ \hline
$n$             & Number of users. \\ 
$[a,b]$         & The range of users' private real-valued data. \\ 
$k$             & Number of samples contained in the perturbed dataset returned by each user. \\ 
$\bm{X}$        & The private real-valued dataset of all users. \\ 
$x_i$           & The private real-valued data of the user $i$. \\
$\bm{X}^{infer}$& The private data of all users obtained by inference. \\
$x^{infer}_i$   & The private data of the user $i$ obtained by inference. \\
$\bm{Y}$        & The perturbed dataset returned by all users. \\ 
$y_{i, j}$      & The $j$th perturbed data returned by the user $i$. \\ 
$F$             & The distribution of all users' private real-valued data. \\ 
$\hat{F}$       & The estimated distribution of all users' private real-valued data. \\ 
$f$             & The probability density function of all users' private real-valued data. \\ 
$\hat{f}$       & The estimated probability density function of all users' private real-valued data. \\ 
$\bm{G}$        & The distribution of all users' perturbed data. \\ 
$\hat{G}$       & The estimated distribution of all users' perturbed data. \\ 
$g$             & The probability density function of all users' perturbed data. \\ 
$p(x,y)$        & The probability density of a user perturbing his private data $x$ to $y$. \\ 
$\bm{Z}$        & A set of points of interest where probability density needs to be estimated. \\ 
$m$             & The number of points of interest. \\ 
$z_i$           & The $i$th points of interest. \\ 
$\bm{V}$        & The estimated set of probability densities of the original data at all sample points. \\ 
$v_i$           & The estimated probability density of the original data at sample point $z_i$. \\ 
\hline
\end{tabular}
\end{table}

Table \ref{tab:symbols} enumerates the notation frequently utilized in this paper, which will be expanded upon in greater detail in subsequent sections when the specific methodologies are introduced. 
Assuming a scenario with $n$ users, where the range of users' private real-valued data is $[a,b]$, we denote the private real-valued data of each user $i$ as $x_i$.
We postulate that each user possesses a client device capable of autonomously executing a straightforward perturbation algorithm on their original data before uploading it to the server. 
Consequently, the server exclusively receives the perturbed data set, denoted as $\bm{Y}=\{y_{1, 1},y_{1, 2}, ..., y_{1, k}, ..., y_{n, 1}, y_{n, 2}, ..., y_{n, k}\}$, where $y_{i, 1}, y_{i, 2}, ..., y_{i, k}$ represent the perturbed data returned by user $i$. 
Leveraging advanced computational capabilities, the server processes the received data to ascertain the distribution $F$ of all users' private real-valued data, characterized by a probability density function $f$. 
Herein,  $\hat{F}$ and $\hat{f}$ are used to signify the estimated distribution and probability density function, respectively.

Our research endeavors are focused on developing a privacy-preserving data collection method tailored for real-valued data. 
Our proposed scheme aims to meet the following design objectives:

\begin{enumerate}
	\item \textbf{Privacy Preservation}: This aspect pertains to the real-valued data of each user within our data collection method. Specifically, it is crucial that the user-returned results cannot facilitate the precise or approximate determination of the users' true data values. This objective ensures that even with the information provided by the users, their personal and sensitive data remain concealed, thereby safeguarding their privacy rights.
	\item \textbf{Distribution Estimation}: Upon receiving information from users, the collector must be capable of directly inferring the overall distribution of all data. This implies that with a single participation, the server can estimate the probability density value for any given value. This objective facilitates efficient and comprehensive data analysis without necessitating repeated data collection.
	\item \textbf{High Utility}: The acquired data distribution must possess sufficient accuracy, and any subsequent data analysis based on this distribution must yield equally precise results. This objective ensures that the collected data is reliable and useful for drawing accurate conclusions and making informed decisions. By maintaining high utility, our data collection method aims to provide valuable insights while respecting users' privacy concerns.
\end{enumerate}

\subsection{Data Perturbation}
%数据收集算法描述
\begin{algorithm}[!t]
	\caption{RVNS\_Perturbation()}
    \label{alg:perturb}
	\SetAlgoLined % 开启行号显示
	\KwIn{
		 $x_i$: the private real-value data of user $i$;\\ 
		$d$: the bandwidth parameter;\\
		$k$: the number of samples contained in the perturbed data set;\\
		$[a, b]$: the range of users' private real-valued data.
	}
	\KwOut{
		$\bm{y}_i$: the perturbed data set of the user $i$
	}
	
	% 负选算法
	\uIf{$(x_i - a \ge d) \&\& (d- x_i \ge d)$}
	{
		$d_1\leftarrow$ Randomly generate a real number from the range of $[0,d]$\;
	}	
	\uElseIf{$x_i - a < d$}
	{
		$d_1\leftarrow$ Randomly generate a real number from the range of $[0,x_i - a]$\;
	}
	\Else
	{
		$d_1\leftarrow$ Randomly generate a real number from the range of $[0, d- x_i]$\;
	}

	$d_2 = d - d_1$\;
	$\bm{y}_i\leftarrow \emptyset$\;		
	\While{$|\bm{y}_i|<k$}{
		\If{$x_i > (a + d_1) \textbf{ AND } x_i < (b - d_2)$}{
			$y\leftarrow$ Randomly sample a real number from the range of $[a,x_i - d_1) \cup (x_i + d_2,b]$\;
			
		}   
		\If{$x_i = a + d_1$}{
			$y\leftarrow$ Randomly select a real number from $(a + d,b]$\;
			
		}
		\If{$x_i = b - d_2$}{
			$y\leftarrow$ Randomly select a real number from $[a,b - d)$\;
			
		}
		$\bm{y}_i \leftarrow \bm{y}_i \cup \{y\}$\;
	}
	\Return $\bm{y}_i$\;
\end{algorithm}

In the realm of categorical sensitive data, the concept of the negative survey model is concerned with the deliberate alteration of the actual category assigned to each user into a negative category, distinctly different from the true category, as delineated in prior works \cite{esponda2009surveys,jiang2017reconstruction}.
Similarly, when considering real-valued data that pertains to user privacy, an augmented level of privacy preservation can be accomplished by introducing perturbations that render the data indistinguishable from its original form. 
Driven by this fundamental principle, we introduce a Real-Value Negative Survey (RVNS) model, specifically designed to perturb users' real-valued privacy data.

Algorithm \ref{alg:perturb} delineates the essential procedures involved in the data perturbation mechanism within the RVNS framework. 
Initially, the RVNS model establishes a range that encompasses the user's private real-valued data and ensures that the selection of perturbed data does not fall within this designated range. 
More precisely, the algorithm generates a real number $d_1$ within the interval $[0,d]$.
Here, $d$ represents a bandwidth parameter, which should be less than $(b-a)$.
Utilizing $d_1$, the prohibited range is defined as $[x_i-d_1, x_i+d-d_1]$, ensuring that its size remains constant at $d$ (Lines 1-2). 
This necessitates that $(x_i-d_1)$ is greater than $a$ and $(x_i+d-d_1)$ is less than $b$, namely $(x_i - a \ge d) \&\& (d- x_i \ge d)$, to maintain the consistency of the prohibited range's size.
Therefore, when $(x_i - a) < d$ or $(d- x_i) > d$, the $d_1$ needs to be generated within the intervals $[0,x_i - a]$ and $[0, d- x_i]$ separately (Lines 3-7).

Subsequently, the RVNS model randomly selects $k$ perturbed data points from the range of private real-valued data $[a,b]$, excluding the prohibited range $[x_i-d_1, x_i+d-d_1]$. 
The sampling strategy employed varies depending on the position of the prohibited range within the overall range $[a,b]$. 
If the prohibited range is centered within $[a, b]$, the perturbed data are sampled from two distinct ranges, including $[a,x_i - d_1)$ and  $(x_i + d_2,b]$ (Lines 11-13).
If the prohibited range is situated at the left boundary of $[a, b]$, the perturbed data are sampled from $(a + d,b]$ (Lines 14-16).
Conversely, if the prohibited range is positioned at the right boundary, the perturbed data are sampled from $[a,b - d)$ (Lines 17-19).

\subsection{Distribution Reconstruction}
From the above data perturbation process, it can be found that the probability density function of original data $f$ and the probability density function of perturbed data $g$ satisfy the following equation.
\begin{equation}
g(y)=\int^a_b p(x,y)\cdot f(x) dx,
\label{equ:gf}
\end{equation}
where $p(x,y)$ is the probability density function of a user with privacy data $x$ perturbing $x$ to $y$.

According to the steps of data perturbation, the probability density function $p(x,y)$ can be divided into three cases.
Due to the page limitation, the derivation details of $p(x,y)$ are provided in the supplementary material.
The first case is that the prohibited range is situated at the left boundary of $[a, b]$, namely $a \le x < a+d$.
According to the range from which the perturbed data are sampled, the $p(x,y)$ can be calculated as follows.
\begin{equation}
\label{equ:pxy_1}
p(x, y) =
\begin{cases}
0 & y \in [x,a+d] \\
\frac{1}{b - a - d} 
& y \in (x+d,b]\\
\frac{x-y}{x-a} \cdot \frac{1}{b-a-d}
& y \in [a,x)\\
\frac{y-a-d}{x-a} \cdot \frac{1}{b-a-d}
& y \in (a+d,x+d]
\end{cases}
\end{equation}

The second case is that the prohibited range is centered within $[a, b]$, namely $a+d\le x \le b-d$.
Since the perturbed data are sampled from $[a,x_i - d_1)$ and  $(x_i + d_2,b]$ in this case, the $p(x,y)$ can be calculated as follows.
\begin{equation}
\label{equ:pxy_2}
p(x, y) =
\begin{cases}
0 & y = x \\
\frac{1}{b - a - d} 
& y \in [a,x-d)\cup (x+d,b]\\
\frac{x-y}{d} \cdot \frac{1}{b-a-d}
& y \in [x-d,x)\\
\frac{y-x}{d} \cdot \frac{1}{b-a-d}
& y \in (x,x+d]
\end{cases}
\end{equation}

In the third case, $x$ belongs to $(b-d,b]$. 
The perturbed data are only sampled from $[a,b - d)$, so the $p(x,y)$ can be calculated as follows.
\begin{equation}
\label{equ:pxy_3}
p(x, y) =
\begin{cases}
0 & y \in [b-d,x] \\
\frac{1}{b - a - d} 
& y \in [a,x-d)\\
\frac{b-d-y}{b-x} \cdot \frac{1}{b-a-d}
& y \in [x-d,b-d)\\  
\frac{y-x}{b-x} \cdot \frac{1}{b-a-d}
& y \in (x,b]
\end{cases}
\end{equation}

Although the Eq. (\ref{equ:gf}) gives the relationship between the probability density function of original and perturbed data, the probability density function of perturbed data is unknown. 
However, there is a group of samples of perturbed data, namely $\bm{Y}$, based on which the probability density function of perturbed data can be estimated by using certain statistical methods.
Considering there might be no prior knowledge about original and perturbed data, the classic non-parametric estimation method, called Kernel Density Estimation (KDE for short) \cite{rosenblatt1956remarks,parzen1962estimation}, is employed to obtain the probability density function of perturbed data.
For better understanding, here we introduce the KDE briefly. 

Assume that there is a group of samples of perturbed data $\bm{Y}$. 
For any point $z\in [a,b]$, KDE uses the following equation to estimate the probability density at point $z$.

\begin{equation}
\label{equ:KDE}
q(z)=\frac{1}{|\bm{Y}|}\sum_{y\in \bm{Y}}\frac{1}{h}K(\frac{z-y}{h}),
\end{equation}
where $|\bm{Y}|$ is the number of samples in $\bm{Y}$, $K(\cdot)$ is the kernel function, and $h\in (0, +\infty )$ is the bandwidth.
There are many kernel functions available for KDE such as uniform kernel, triangular kernel, and Gaussian kernel. 
Due to the smoothness assumption, the Gaussian kernel is a widely used kernel function. 
Therefore, in this paper, we adopt the Gaussian kernel as the kernel function, which is given as follows.
\begin{equation}
\label{equ:gauss}
K(t)=\frac{1}{\sqrt{2\pi}}e^{-\frac{t^2}{2}}
\end{equation}
Using Eqs. (\ref{equ:KDE}) and (\ref{equ:gauss}), the probability density of any point can be estimated from perturbed data returned by the users.

However, the KDE can only obtain the probability density values at different points rather than the formal expression of the probability density.
Fortunately, there are many numerical methods for calculating definite integrals based on sampling points, such as rectangular rule \cite{powell1985simpson}, trapezoidal rule \cite{abdulhameed2021improved}, Romberg integration \cite{tseng2008digital}, and so on.
Using either numerical method, the relationship between the probability density of the original data and the perturbed data can be approximated at a set of sample points.
Considering the simplicity, the rectangular rule is employed to approximate Eqs. (\ref{equ:gf}), which is given as follows.
\begin{equation}
g(z)=\sum_{z_i\in \bm{Z}} p(z_i,z)\cdot f(z_i)\cdot (z_{i+1}-z_i),
\label{equ:gf-app}
\end{equation}
where $\bm{Z}=(z_1, z_2, ..., z_m)$ is a set of points of interest of which probability density needs to be estimated, and $z$ is either of them.
Here, the elements in $\bm{Z}$ are sorted from small to large, and an additional sampling point $z_{m+1}>z_m$ is added for the convenience of calculation, of which probability density is not required.

Given $m$ samples points $\bm{Z}$, Eq. (\ref{equ:gf-app}) determines $m$ equations with $m$ variables, which are given as follows.
\begin{equation}
\label{equ:opt-group}
G(\bm{Z})=\bm{P}(\bm{Z})\cdot F(\bm{Z})^T,
\end{equation}
where $G(\bm{Z})=(g(z_1), g(z_2), ..., g(z_m))$, $\bm{P}(\bm{Z})=(p(z_1,z_1), p(z_2,z_1), ..., p(z_m,z_1), p(z_1,z_2), ..., p(z_m,z_m))$, $F(\bm{Z})=(f(z_1), f(z_2), ..., f(z_m))$, and $F(\bm{Z})^T$ is the transpose of $F(\bm{Z})$.
By using matrix inversion \cite{krishnamoorthy2013matrix} or maximum likelihood estimation \cite{pan2002maximum}, the probability density of the original data $F(\bm{Z})$, namely the distribution of original data, can be obtained from Eq. (\ref{equ:opt-group}).
However, in some cases, the matrix $\bm{P}(\bm{Z})$ is irreversible, leading to the infeasibility of the matrix inversion-based method.
As for the maximum likelihood estimation, due to it obtaining probability density of the original data from the perspective of expectation maximization without considering the overall distribution similarity, the estimated distribution accuracy obtained is limited.

To this end, a reconstruction method considering distribution similarity is developed to estimate the probability density of the original data from the perturbed data.
Let's turn our attention back to Eq. (\ref{equ:gf-app}).
Eq. (\ref{equ:gf-app}) gives the relationship between the probability density of the original data $F(\bm{Z})$ and the perturbed data $G(\bm{Z})$.
For a given estimated probability density of the original data $\hat{F}(\bm{Z})$, Eq. (\ref{equ:gf-app}) can give its corresponding probability density of the perturbed data $\hat{G}(\bm{Z})$.
Since the probability density of the perturbed data of original data $G(\bm{Z})$ estimated by the KDE is known, we can use the similarity between $\hat{G}(\bm{Z})$ and $G(\bm{Z})$ to measure the similarity between $\hat{F}(\bm{Z})$ and $F(\bm{Z})$.
Based on the above idea, the following optimization problem is built to help estimate the probability density of the original data.
{
\begin{eqnarray} 
%\label{equ:opt-KL}
\textrm{min}\ F_{obj}(\bm{V})=&&KL(G(\bm{Z}),\bm{P}(\bm{Z})*\bm{V}^T) \nonumber\\
&&+ KL(\bm{P}(\bm{Z})*\bm{V}^T,G(\bm{Z}))\nonumber\\
&&+L_1(\bm{V})+L_2(\bm{V}) \label{equ:obj}\\
\textrm{s.t.} &&\sum_{v_i\in \bm{V}} v_i\cdot (z_{i+1}-z_i)=1 \label{equ:con1}\\
&& 0\le v_i \le 1, \forall v_i\in \bm{V} \label{equ:con2}
\end{eqnarray}
}
In the above optimization problem, $\bm{V}=(v_1, v_2, ..., v_m)$ is the decision vector, $v_i$ represents the estimated probability density of the original data at sample point $z_i$, namely $f(z_i)$, $KL(G(\bm{Z}),\bm{P}(\bm{Z})*\bm{V}^T)$ represents the Kullback-Leibler divergence \cite{li2020estimating} between $G(\bm{Z})$ and $\bm{P}(\bm{Z})*\bm{V}^T)$, $L_1(\bm{V})$ is the L1 regularization of $\bm{V}$, and $L_2(\bm{V})$ is the L2 regularization of $\bm{V}$.
Eq. (\ref{equ:obj}) gives the objective of the problem, which measures the distribution similarity between the probability density of perturbed data $G(\bm{Z})$ and the probability density of perturbed data corresponding to $\bm{V}$.
Here, L1 and L2 regularization is used to ensure the smoothness of estimated distribution and prevent it from overfitting to points with lower probability densities \cite{yang2023deep}.
Eq. (\ref{equ:con1}) is the area constraint, which ensures the integral of the estimated probability density over the range $[a, b]$ is equal to 1.
Eq. (\ref{equ:con2}) is the bound constraint, which limits the probability density of each point between 0 and 1.
By solving the above problem, we can reconstruct the distribution of original data from the perspective of distribution similarity.
In this paper, Sequential Quadratic Programming (SQP) \cite{gill2011sequential} is employed to solve the above problem, owing to the nonlinear characteristics of Kullback-Leibler divergence.

\begin{algorithm}[t]
	\caption{RVNS\_Reconstruction()}
    \label{alg:esti}
	\SetAlgoLined % 开启行号显示
	\KwIn{$\bm{Y}=\{y_{1, 1},y_{1, 2}, ..., y_{1, k}, ..., y_{n, 1}, y_{n, 2}, ..., y_{n, k}\}$: the perturbed data returned by all users; $\bm{Z}=(z_1, z_2, ..., z_m)$: a set of points of interest of which probability density needs to be estimated.
	}
	\KwOut{
		$\hat{F}(\bm{Z})=(\hat{f}(z_1), \hat{f}(z_2), ..., \hat{f}(z_m))$: the estimated values of probability density at points in $\bm{Z}$.
	}
	
	% 重构算法
	$\bm{P}(\bm{Z}) \leftarrow$ Calculate element values in $\bm{P}(\bm{Z})$ by using Eqs. (\ref{equ:pxy_1})-(\ref{equ:pxy_3})\;
	$\bm{G}(\bm{Z}) \leftarrow$ Estimate the probability density of perturbed data at points in $\bm{Z}$ based on $\bm{Y}$ by using the KDE, namely Eq. (\ref{equ:KDE})\;
	$\hat{F}(\bm{Z}) \leftarrow$ Solve optimization problem defined by (\ref{equ:obj})-(\ref{equ:con2}) using the SQP \;
	\Return $\hat{F}(\bm{Z})$\;
\end{algorithm}

On the whole, the main steps of the RVNS reconstructing distribution are summarized in Algorithm \ref{alg:esti}.
Specifically, the RVNS first determines the matrix $\bm{P}(\bm{Z})$ according to Eqs. (\ref{equ:pxy_1})-(\ref{equ:pxy_3}) (Line 1).
Then, the probability density of perturbed data at points in $\bm{Z}$, namely $\bm{G}$, is estimated from $\bm{Y}$ by using the KDE (Line 2).
Finally, the optimization problem (\ref{equ:obj})-(\ref{equ:con2}) is solved by using the SQP to reconstruct the probability density of original data at points in $\bm{Z}$ (Line 3).

\section{Theoretical Analysis}
\label{sec:analy}
As indicated in the above section, the developed scheme should meet the three design objectives, including privacy preservation, distribution estimation, and high utility.
Therefore, this section analyzes the RVNS from the above three aspects.
Moreover, the time complexity of the RVNS is also analyzed at the end of this section.

\subsection{Privacy Preservation}
The privacy-preserving ability of the RVNS is ensured by the novel local differential privacy model \cite{yang2023local}, of which the definition is given as follows.

\begin{defi}[$\epsilon$-Local Differential Privacy]
	A randomized mechanism $\mathscr{M}$ satisfies $\epsilon$-local differential privacy if and only if for any two inputs $x_1$ and $x_2$ in the domain of $\mathscr{M}$, and for any possible output $y$ of $\mathscr{M}$, the following condition holds:
	\begin{equation} 
	Pr[\mathscr{M}(x_1) = y] \le e^{\epsilon} \cdot Pr[\mathscr{M}(x_2) = y],
	\end{equation}
	where $Pr[\cdot]$ denotes the probability and $\epsilon$ is the privacy budget.	
\end{defi}

However, the output of the RVNS is continuous, so strictly speaking, the probability of the RVNS outputting any specific value is zero.
In practical applications, when two output values are extremely close, they can be considered the same.
Therefore, we can select a small number $\delta$ to determine a neighborhood of a value and consider the output that falls within a certain value neighborhood to be the same as this value.
In other words, we use the probability of outputting a value that falls within $[y-\delta, y+\delta]$ to approximate the probability of the RVNS outputting $y$.
\begin{theo}
	\label{theo1}
	The RVNS satisfies $k\ln(\frac{4d}{\delta})$-local differential privacy.
\end{theo}

\begin{proof}  
	In the case of $k=1$, for any input values $x_1, x_2 \in [a,b]$ and any possible output $y\in [a,b]$, we have
	\begin{equation} 
	\label{equ:proof}
	\frac{Pr[\mathscr{M}(x_1) = y]}{Pr[\mathscr{M}(x_2) = y]}=\frac{\int_{y-\delta}^{y+\delta} p(x_1,y) dy}{\int_{y-\delta}^{y+\delta} p(x_2,y) dy}
	\end{equation}
	According to the definition of $p(x,y)$, the above formulation can reach its maximum value when $|x_1-y|>d$, and $(a\le x_2 < a+d)\&\& (y-\delta<a+d<y+\delta \le x_2+d)$.
	Therefore, we have
	\begin{equation}
    \begin{aligned}
    \frac{Pr[\mathscr{M}(x_1) = y]}{Pr[\mathscr{M}(x_2) = y]} &\le \frac{\int_{y-\delta}^{y+\delta} \frac{1}{b - a - d} \, dy}{\int_{a+d}^{y+\delta} \frac{y-a-d}{x_2-a} \cdot \frac{1}{b-a-d} \, dy} \\
    &= \frac{4\delta(x_2-a)}{(x_2+\delta-a-d)^2}
    \end{aligned}
    \end{equation}
	Since $a\le x_2 < a+d$, we have
	\begin{equation}
	\frac{Pr[\mathscr{M}(x_1) = y]}{Pr[\mathscr{M}(x_2) = y]}\le \frac{4d}{\delta}
	\end{equation}
	Therefore, the RVNS satisfies $\ln(\frac{4d}{\delta})$-local differential privacy, when $k=1$.
	Further, when $k>1$, the RVNS is essentially equivalent to executing $k$ times the algorithm set in the case of $k=1$.
	Consequently, according to the sequential composition theorem of differential privacy, the RVNS satisfies $k\ln(\frac{4d}{\delta})$-local differential privacy, when $k>1$.			
\end{proof}  

Although when $\delta$ tends to infinitely small, $k\ln(\frac{4d}{\delta})$ tends to infinitely great, in practical applications, $\delta$ should not be set too small because a too small $\delta$ will assume that the adversary has not obtained the original data even if he has inferred a very close value.
Therefore, the RVNS satisfies the local differential privacy model, and can effectively protect user privacy in practical applications, even if the perturbed data has already been known by the adversary.

Moreover, from the process of RVNS perturbing the data, it can be found that the RVNS will never output values that are the same as the original data.
Even if the output that falls within a certain value neighborhood is the same as this value, the probability of the RVNS outputting a value that belongs to the neighborhood of the original data is much lower than that of other data.
Therefore, compared to existing methods, the RVNS can make users feel more at ease, thereby improving their level of cooperation.

\subsection{Distribution Estimation}
The RVNS can estimate the probability density value of any point belonging to the range of $[a, b]$ based on the results returned by all users participating once.
From the steps of reconstruction shown in Algorithm \ref{alg:esti}, it can be found that the reconstruction process of the RVNS only requires two inputs, including the perturbed data returned by all users $\bm{Y}$ and a set of points of interest $\bm{Z}$.
The inputs $\bm{Y}$ and $\bm{Z}$ are independent of each other.
The value of $\bm{Z}$ can be set according to actual needs.
By contrast, in the methods that need prior discretization of the data, the user needs to return the data once the set of points of interest at which the probability density is to be estimated is changed.
Consequently, compared to the methods that require prior discretization of the data, the RVNS can obtain the probability density value of any point more efficiently.

\subsection{High Utility}
The RVNS ensures the high utility of estimated distribution from the following two aspects.
On the one hand, the RVNS employs the KDE to directly estimate the probability density of perturbed data and does not require the discretization of the data.
Therefore, it does not own the disadvantage of ignoring the difference of values within the same bin.
In contrast, the methods that require posterior discretization of the data still need to divide the data into different bins, which overlook the intrinsic differences among data within the same bin, and restrict the utility of estimated distribution.
On the other hand, the RVNS models the reconstruction process as an optimization problem from the point of view of distribution similarity, to ensure the similarity between the obtained distribution and the original distribution.
However, existing methods estimate the probability density of different points in a relatively independent process, which only considers that the whole range of the probability density integral result is equal to one, so the accuracy of the estimated distribution is limited.
Therefore, compared to existing methods the RVNS can obtain more accurate distribution estimation results.

\subsection{Complexity Analysis}
On the user side, the RVNS holds the time complexity of $\bm{O}(k)$, where $k$ is the number of perturbed data returned by each user.
From the Algorithm \ref{alg:perturb}, it can be found that on the user side, the RVNS contains two steps, including determining prohibiting range and sampling perturbed data.
The time complexity of the first steps is $\bm{O}(1)$ because it only requires to generate a random number.
The second step has a time complexity of $\bm{O}(k)$, given that $k$ random numbers are required to determine the perturbed data.
Overall, the time complexity of the RVNS on the user side is $\bm{O}(k)$.

On the server side, the RVNS owns the time complexity of $\bm{O}(m\cdot k\cdot n + T\cdot m^3)$, where $T$ is the number of iterations for SQP and $m$ is the number of points of interest.
From the Algorithm \ref{alg:esti}, it can be found that on the server side, the RVNS contains three steps, including calculating matrix $\bm{P}$, estimating probability density, and estimating distribution.
The time complexity of calculating the matrix is $\bm{O}(m^2)$, because the scale of the matrix is $m\times m$, and the value of each element can be directly calculated.
The second step has a time complexity of $\bm{O}(m\cdot k\cdot n)$.
This is because the probability density calculation of each point of interest requires traversing all the perturbed data returned by users.
As for the third steps, its time complexity is $\bm{O}(T\cdot m^3)$, given that the SQP requires $\bm{O}(m^3)$ time complexity for each iteration \cite{boggs1995sequential}.
Overall, the time complexity of the RVNS on the server side is $\bm{O}(m\cdot k\cdot n + T\cdot m^3)$.

Based on the above analysis, it can be found that the RVNS can not only obtain high-utility estimated distributions while protecting users' personal privacy but also have high efficiency.

\section{Experimental studies}
\label{sec:exper}
\subsection{Compared Algorithms and Data Sets}

\subsubsection{Compared Algorithms}
To evaluate the performance of the proposed RVNS, we compare it against several state-of-the-art perturbation-based privacy preservation approaches, including Laplace-DP \cite{gough2021preserving}, Gaussian-pDP \cite{liu2018generalized}, Flipped Huber \cite{10223238hunhe}, and EMS \cite{li2020estimating}.
The Laplace-DP employs Laplacian noise addition for differential privacy to protect user data, particularly in the context of electricity consumption information.
The Gaussian-pDP generalizes the widely used Laplacian mechanism to the family of Generalized Gaussian (GG) mechanisms to safeguard user privacy.
The Flipped Huber introduces a novel differential privacy noise addition mechanism, where noise is sampled from a mixed density resembling a centered Laplacian and tail-heavy Gaussian, significantly enhancing privacy preservation compared to traditional methods.
As for the EMS, it leverages random response technology to perturb original sensitive data and proposes a Smoothed Expectation Maximization (EMS) algorithm to reconstruct the processed data, utilizing aggregated histograms to estimate the original data distribution.

All methods were implemented using MATLAB. 
Experiments were conducted on a system running MATLAB R2021b with a 12th Gen Intel(R) Core(TM) i5-12400 processor at 2.50 GHz and 24 GB of RAM. 
For each method, experiments were repeated 11 times, and the average results were reported to ensure reliability and consistency.
Due to the page limitation, the sensitivity of parameters in our RVNS is analyzed in the supplementary material.

%and in addition, the combination of NetNS and PBCN\cite{huang2020privacy} with four data publishing methods(i.e., $k$-anonymity\cite{WOS:000179217500008}, $l$-diversity \cite{machanavajjhala2007diversity}, $t$-closeness \cite{WOS:000277717500003},KAB \cite{WOS:000827405000015}) is used to verify the effectiveness of our proposed method NetNR on the privacy preservation of node attributes in attribute social networks and on the privacy preservation of topology privacy preservation in attribute social networks. The PBCN randomly disturbs the edges in the social network to satisfy the differential privacy model. 

%The data publishing methods, i.e., $k$-anonymity, $l$-diversity, $t$-closeness, are the most widely used generalization methods for data privacy preservation, which $k$-anonymity first divides the original data into multiple groups then generalizes each group of records so that they have the same quasi-identifier values, $l$-diversity guarantee at least $l$ values of sensitive attributes for each equivalence class on the basis of $k$-anonymity, and $t$-closeness makes the distribution of data after $k$-anonymity must not differ much from the original data by $Mean$s of probability distribution. Moreover, the KAB clusters the data into $k$ groups based on the clustering method and uses the black hole algorithm to find the optimal solution and then generalizes the data.

\subsubsection{Tested DataSets}
In this study, we employ a comprehensive collection of datasets, consisting of five synthetic datasets and three real-world datasets, all composed of numerical values. 
The synthetic datasets are generated from Chi-squared distributions with degrees of freedom (df) of two, three, five, eight, and ten, respectively. 
Each synthetic dataset comprises 50,000 data points, ranging within $[0,10]$, namely $a=0$, and $b=10$.
For each synthetic dataset, 100 sampling points are uniformly sampled within its range of values to estimate the probability density of the data points.

The real-world datasets include BMI \cite{BMI}, Bagrut Grade \cite{grade}, and IMDB Rating \cite{Movies}.
The BMI dataset contains Body Mass Index (BMI) data from patients. 
Initially containing 68,205 samples within the range $[3.4718, 298.6667]$, we filter the dataset to include 68,179 samples within $[0,100]$, and uniformly sample 1,000 within $[0, 100]$ for probability density estimation.
The Bagrut Grade dataset encompasses average Bagrut grades from over 1,800 schools across various subjects from 2013 to 2016. 
Initially containing 69,638 samples, after removing missing values, 54,738 valid samples remained within the range [0, 100]. 
Also, 1,000 sampling points are used for probability density estimation.
IMDB Rating dataset includes IMDB ratings for over 46,000 movies spanning from 1874 to 2016. 
After excluding missing data, 44,300 valid samples remained within the range $[0, 10]$. 
For the IMDB Rating dataset, 100 sampling points are uniformly sampled from $[0, 10]$ for probability density estimation.

\subsection{Privacy and Utility Metrics}
\subsubsection{Privacy Metric}
In the realm of privacy-preserving data analysis, the parameter $\epsilon$ in the local differential privacy has traditionally been employed as a metric to quantify privacy. 
However, in our RVNS, the computation of $\epsilon$ is intricately tied to the neighborhood size defined during the data analysis phase, namely $\delta$. 
To ensure fairness in comparisons across different approaches, we adopt a privacy measure indicator based on Euclidean distance, as suggested in prior literature \cite{park2018data}.
This metric evaluates the privacy preservation capability of a method by assessing the proximity of the perturbed data to the original data, which can be calculated as follows.
\begin{equation}
dis(\bm{X}, \bm{X}') = \sqrt{\sum_{i=1}^{n} (x_i - x'_i)^2},
\end{equation}
where $\bm{X}$ is the set of original data, while $\bm{X}'$ is the set of perturbed data.
A larger Euclidean distance indicates superior privacy preservation.

Given that in our RVNS, a user may return multiple perturbed data, we devise a dedicated method to infer the original data from the perturbed data set generated by our RVNS. 
Specifically, considering an original dataset $\bm{X}=\{x_1, x_2, ..., x_n\}$ of $n$ users, where each user returns $k$ perturbed data samples, resulting in a perturbed dataset $\bm{Y}=\{y_{1, 1},y_{1, 2}, ..., y_{1, k}, ..., y_{n, 1}, y_{n, 2}, ..., y_{n, k}\}$ containing $k\cdot n$ elements. 
Assuming that an adversary possesses knowledge of our perturbing strategy and attempts to reverse-engineer the truly sensitive data of each user $i$ through the perturbing probabilities $p(x,y)$ and his perturbed dataset $\bm{y}_i=\{y_{i, 1},y_{i, 2}, ..., y_{i, k}\}$, this constitutes the following optimization problem.
\begin{eqnarray}
\textrm{min}\ F_{infer} (x^{infer}_i) = && -\prod_{j=1}^{k} p(x^{infer}_i, y_{i, j}) \\
\textrm{s.t.} && a \leq x^{infer}_i \leq b
\end{eqnarray}
The above optimization problem is to infer the original data of user $i$ by determining an $x^{infer}_i$ that has the highest probability of obtaining perturbation data of user $i$.
By applying the above optimization problem to all users, the adversary can infer a dataset $\bm{X}^{infer}=\{x^{infer}_1, x^{infer}_2, ..., x^{infer}_n\}$.
Subsequently, the Euclidean distance between the original sensitive information set $\bm{X}$ and the inferred dataset $\bm{X}^{infer}$ can be used to assess the privacy of our RVNS. 

\begin{figure*}[!t]
	\centering
	\subfloat[$\chi^2(2)$]{\includegraphics[width=0.23\linewidth]{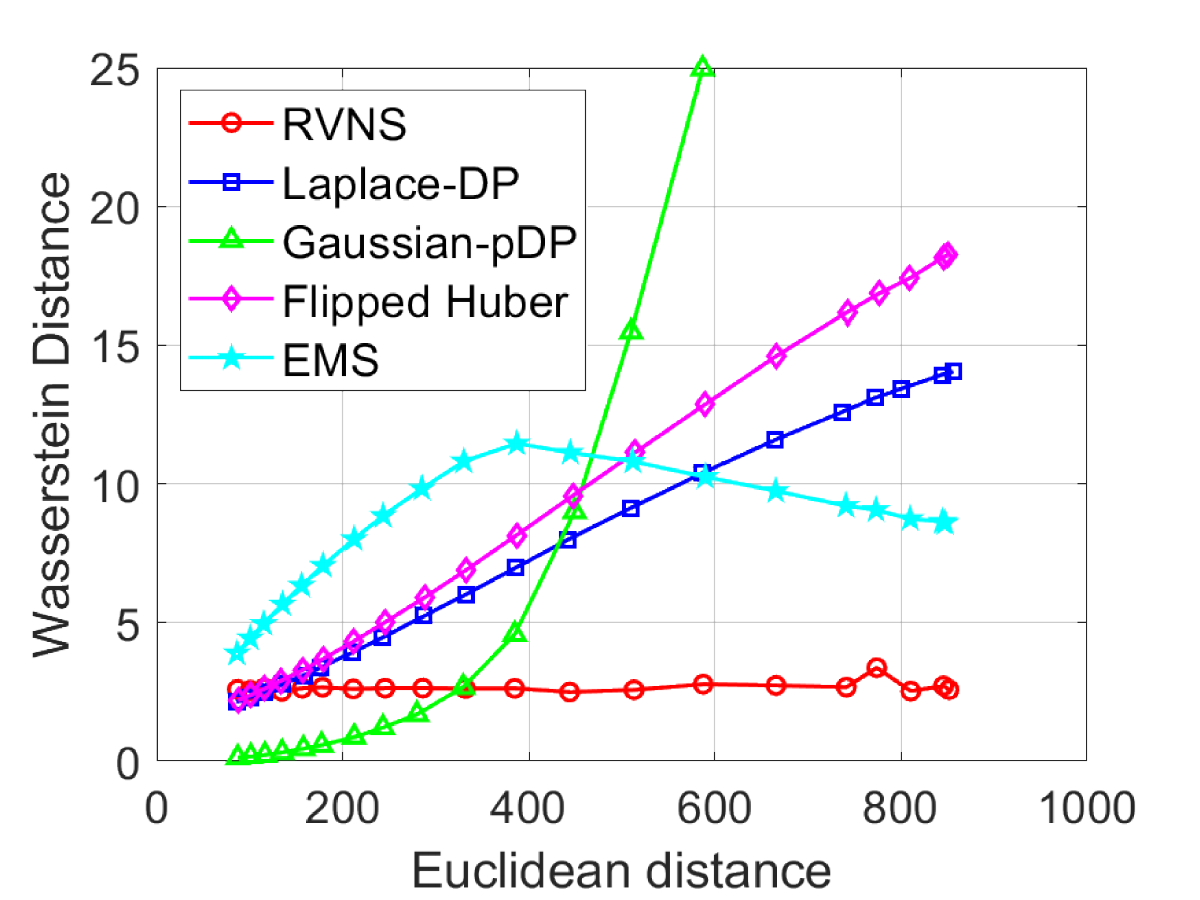}}
	\hfil
	\subfloat[$\chi^2(3)$]{\includegraphics[width=0.23\linewidth]{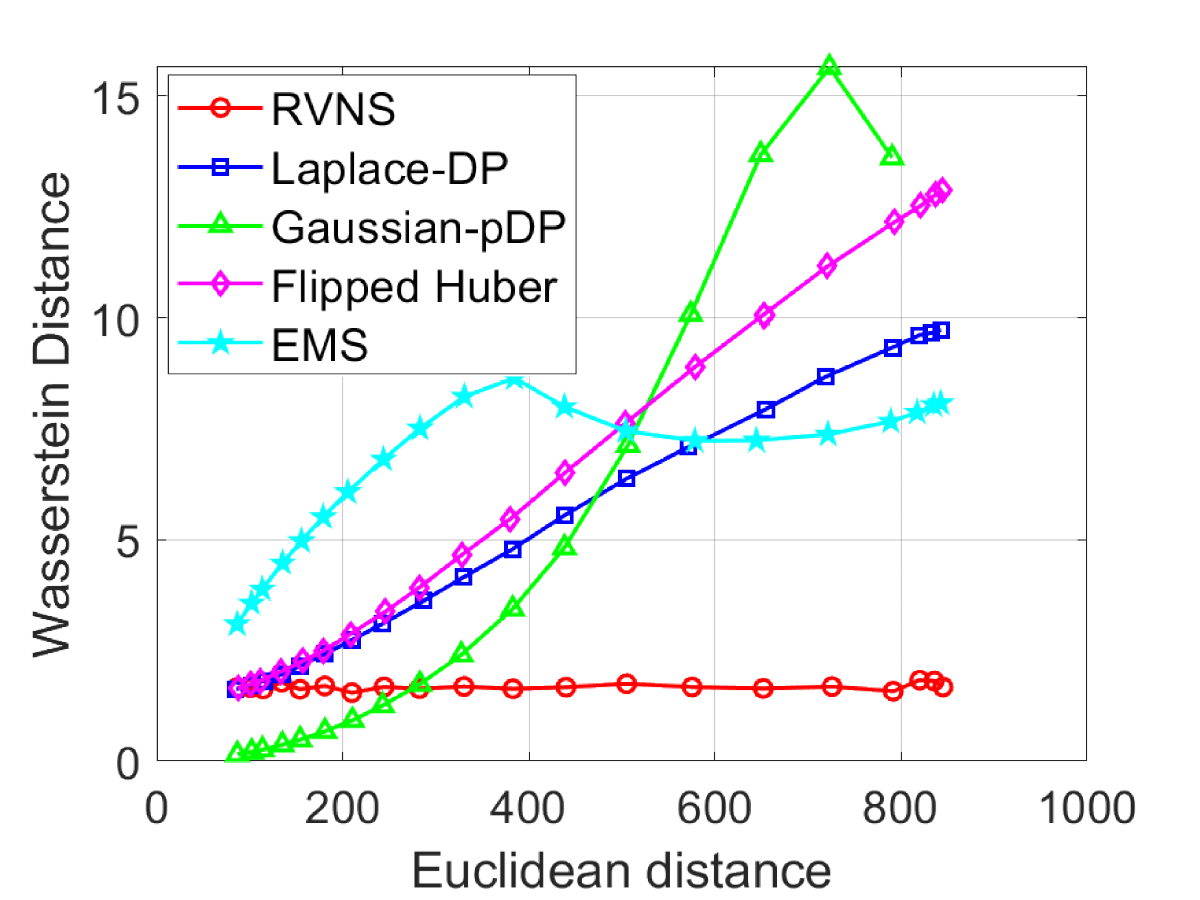}}
	\hfil
	\subfloat[$\chi^2(5)$]{\includegraphics[width=0.23\linewidth]{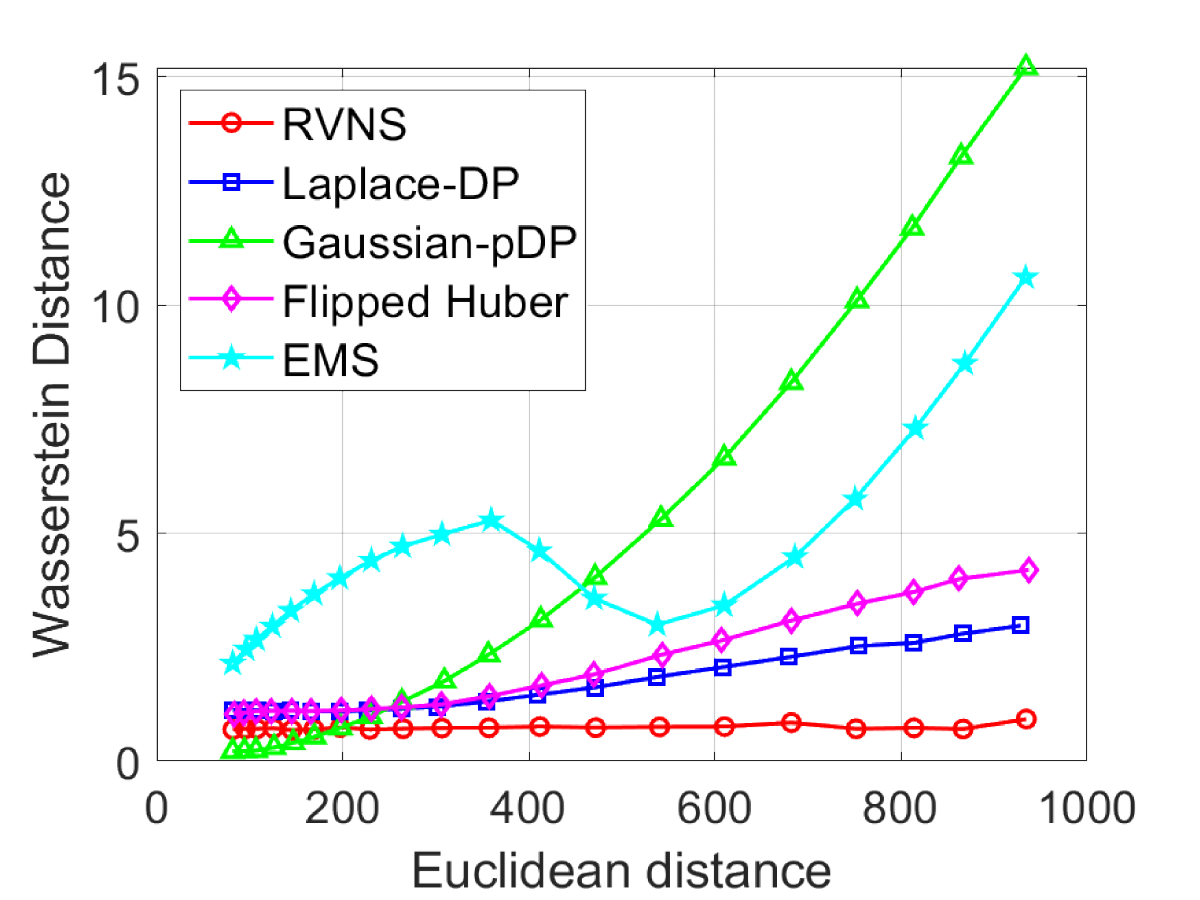}}
	\hfil
	\subfloat[$\chi^2(8)$]{\includegraphics[width=0.23\linewidth]{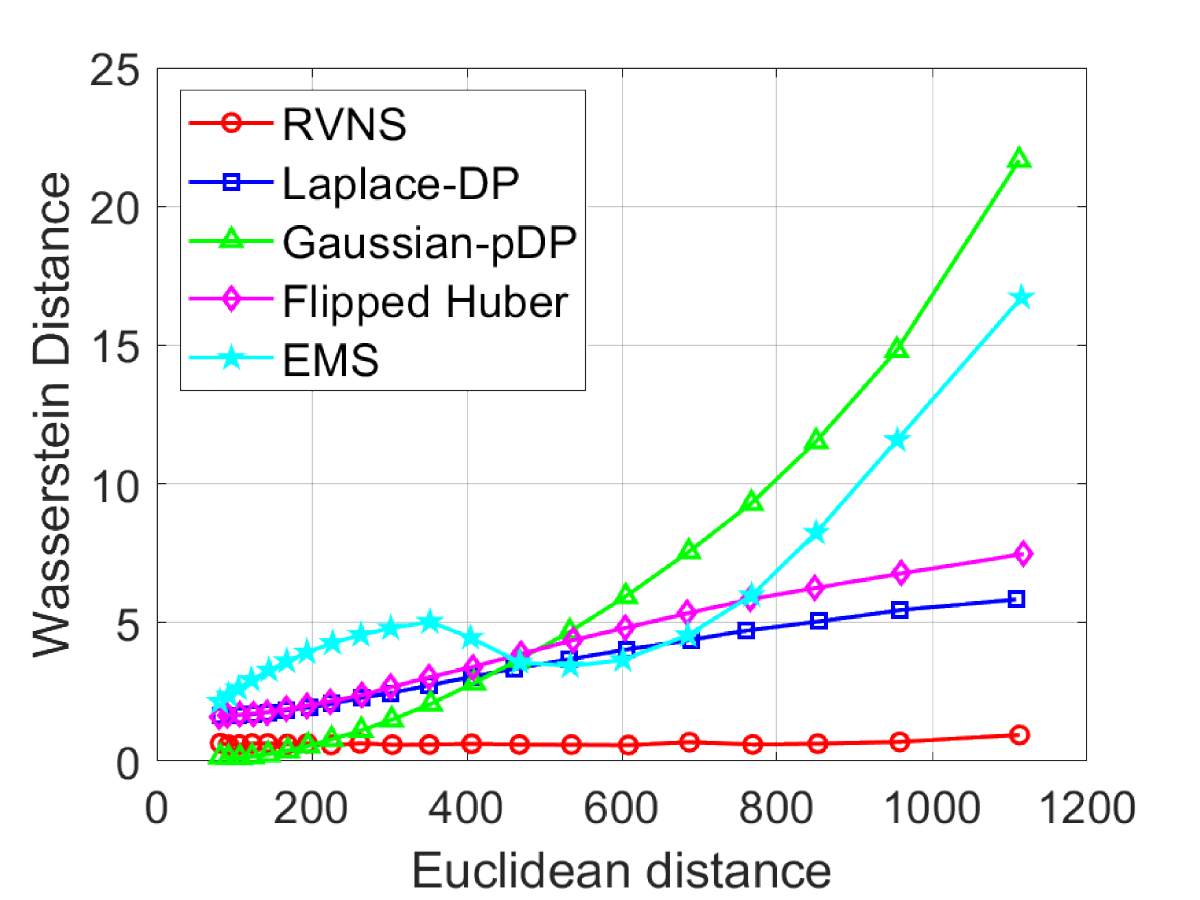}}
	\hfil
	\subfloat[$\chi^2(10)$]{\includegraphics[width=0.23\linewidth]{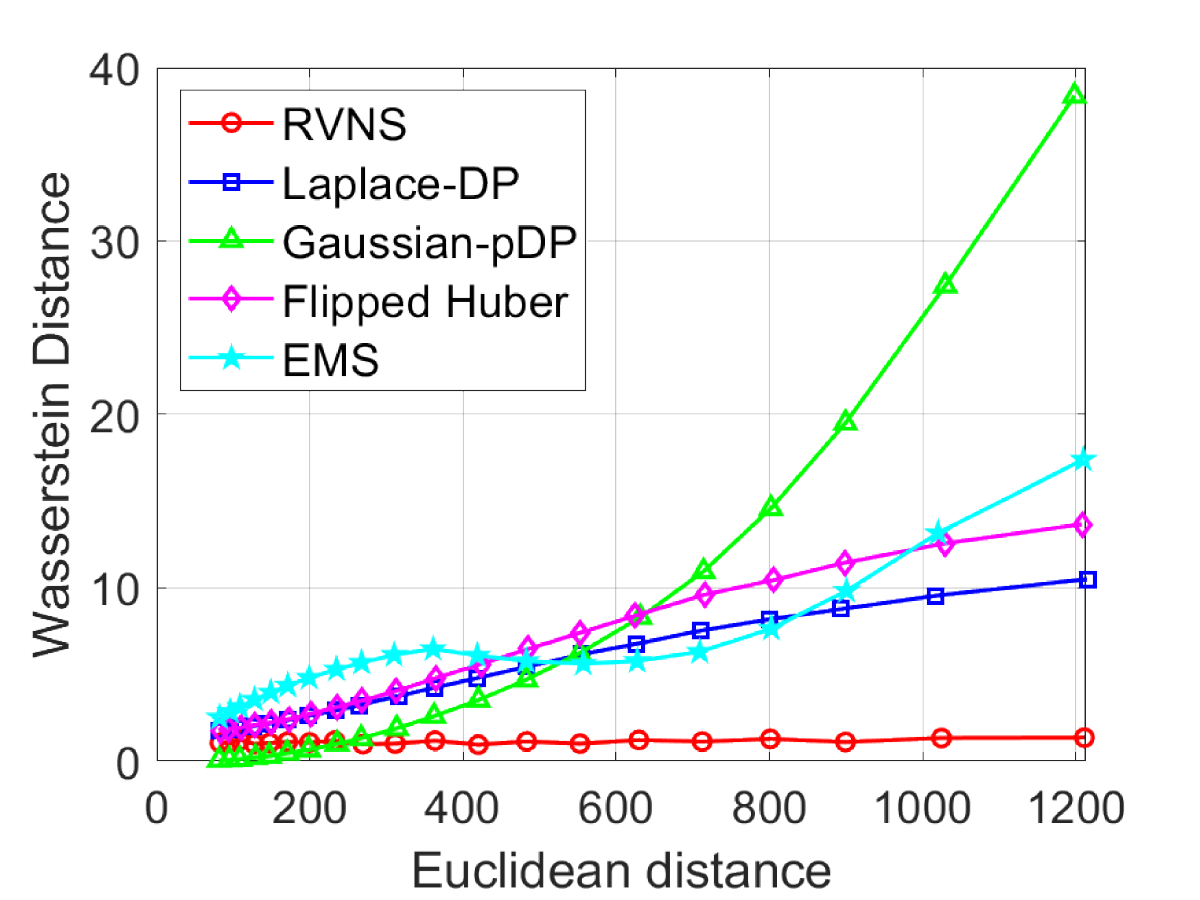}}
	\hfil
	\subfloat[BMI]{\includegraphics[width=0.23\linewidth]{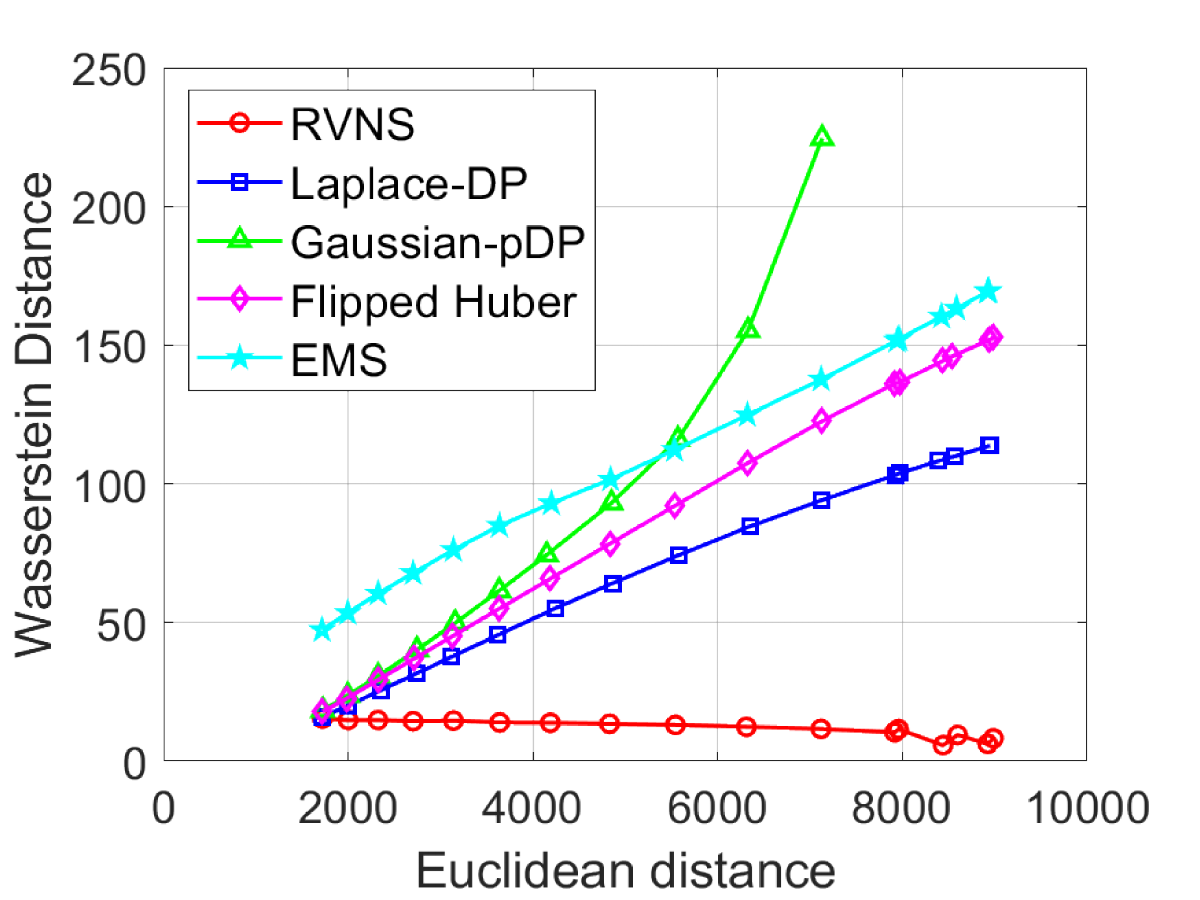}}
	\hfil
	\subfloat[Bagrut Grade]{\includegraphics[width=0.23\linewidth]{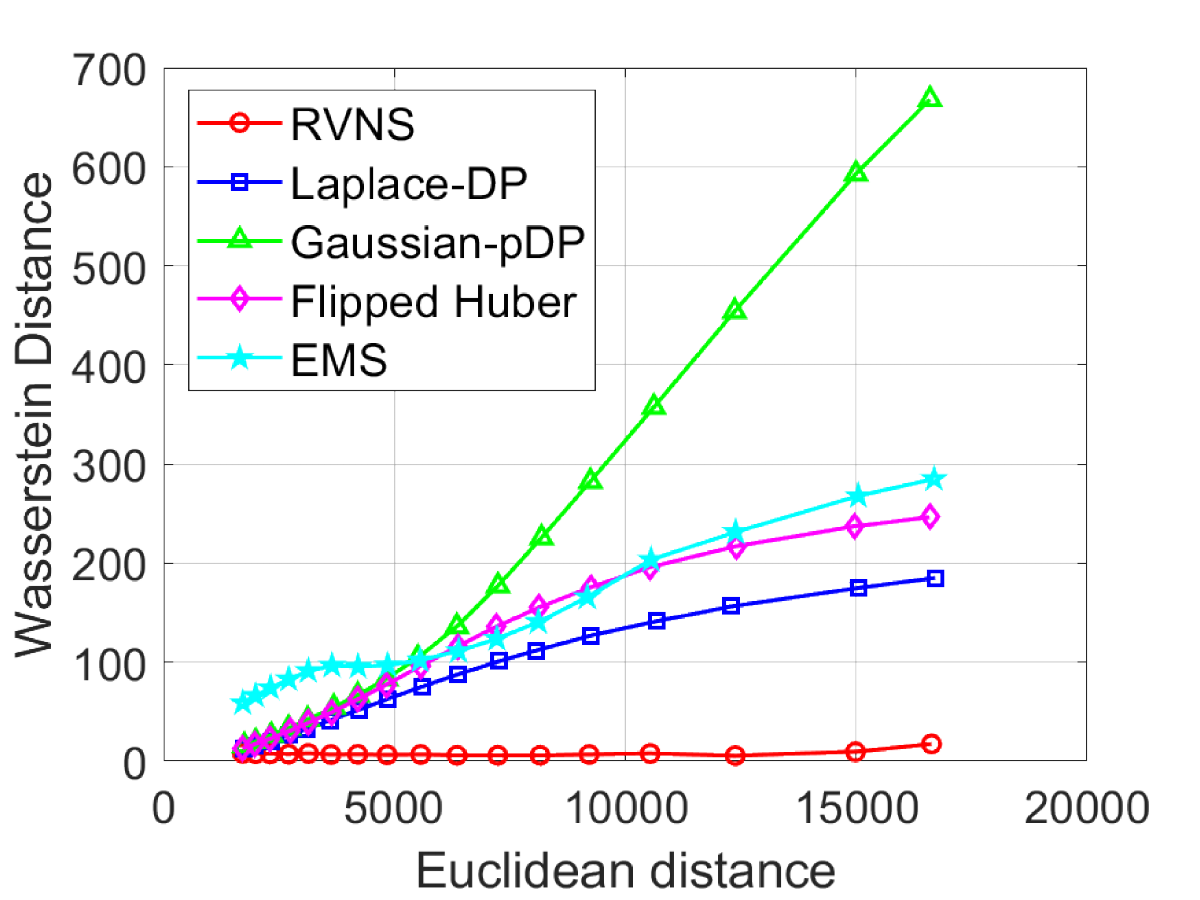}}
	\hfil
	\subfloat[IMDB Rating]{\includegraphics[width=0.23\linewidth]{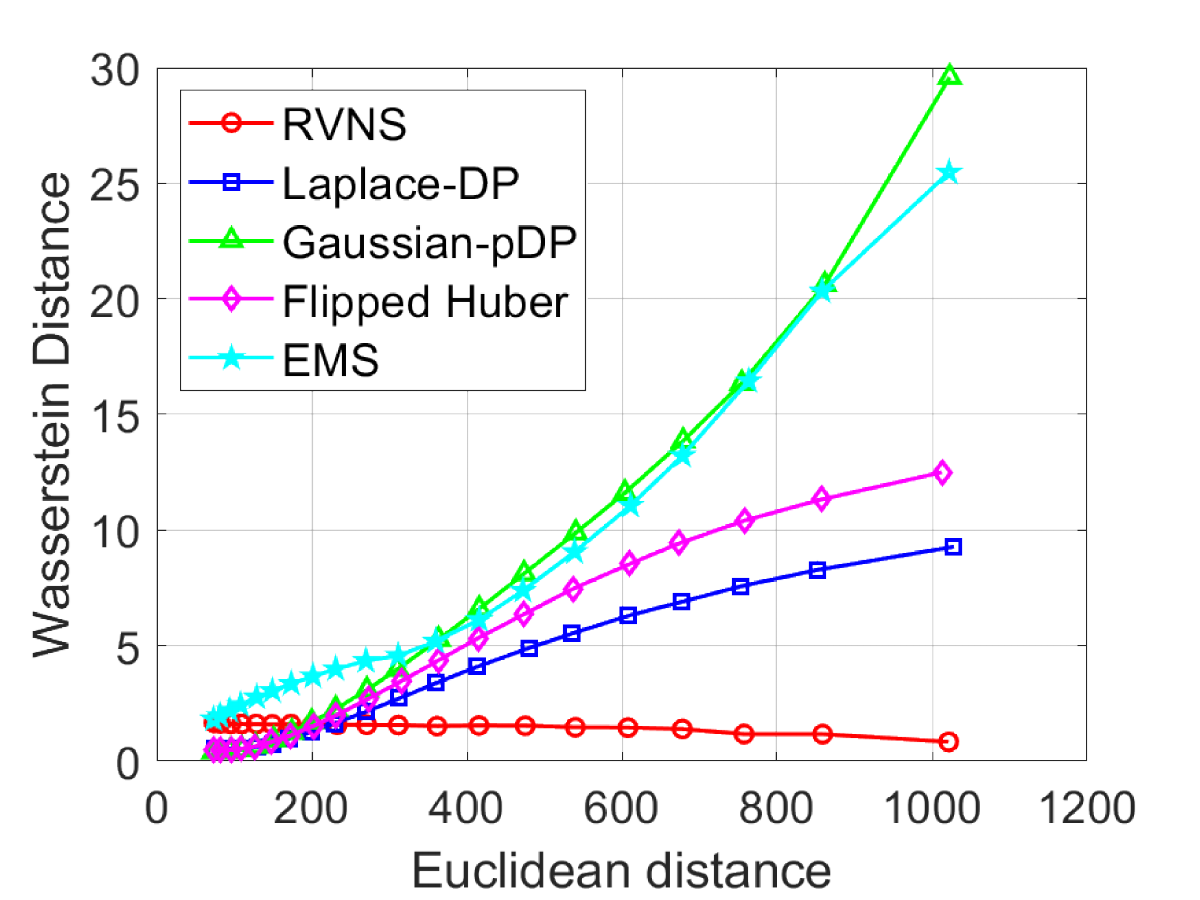}}	
	\caption{The comparison of distribution similarity of data perturbed by the RVNS and four compared algorithms on five synthetic datasets and three real-world datasets under different privacy preservation levels.}
	\label{fig:WE}
\end{figure*}

\subsubsection{Utility Metrics}
In this paper, two different types of metrics are employed to measure the utility of data obtained by different methods.
The first type of metric directly measures the distribution similarity of probability density between original data and data obtained by different methods.
Here, the Wasserstein Distance \cite{li2020estimating}, which represents the minimum cost required to transport the mass of one distribution to match the other, is employed as a pivotal metric for quantifying the discrepancy between two probability distributions. 
For two one-dimensional discrete distributions $R = \{r_i\}_{i=1}^n$ and $S = \{s_i\}_{i=1}^n$, the one-dimensional Wasserstein Distance $W_1(R,S)$ can be calculated as follows. 
\begin{equation}
W_1(R,S) = \sum_{i=1}^{m} |F_R(i) - F_S(i)|,
\end{equation}
where, $F_R(i)= \sum_{j=1}^{i} r_j$ and $F_S(i) =\sum_{j=1}^{i} s_j$ represent the cumulative distribution functions (CDFs) of distributions $R$ and $S$, respectively, while $m$ denotes the number of discrete data points. 
This formulation captures the absolute differences between the CDFs of the two distributions, providing a precise measure of their similarity.

The second type of metrics employs six widely used fundamental statistical indicators \cite{tai2023user} to further validate the usability of collected data across diverse statistical analysis methods.
These six indicators include $Mean$, $Standard\ Deviation$, $Mode$, $Median$, $Skewness$, and $Kurtosis$.
By utilizing these six statistical indicators, we can comprehensively evaluate the similarity between the original data and the data obtained by different methods. 
Specifically, we first estimate the probability density from the data obtained by different methods.
Here, if the method cannot directly output the probability density, the KDE is used to estimate the probability density from the perturbed data.
After that, we sample from the estimated probability density to create a new dataset $\bm{\hat{X}}$ with the same size as the original sensitive dataset $\bm{X}$.
Subsequently, we compute the aforementioned statistical indicators based on dataset $\bm{\hat{X}}$, and then use the absolute value of the difference in statistical indicators values calculated on datasets $\bm{X}$ and $\bm{\hat{X}}$ to assess the utility.

\subsection{Performance Analysis}
\subsubsection{Distribution Similarity}
Fig. \ref{fig:WE} plots the trade-off between the privacy level and the distribution similarity of data perturbed by the RVNS and four compared algorithms on five synthetic datasets and three real-world datasets.
The analysis of the experimental results, presented in Fig. \ref{fig:WE}, yields the following two principal observations.

Firstly, under equivalent levels of privacy preservation, our RVNS achieves overall the best estimates of data distributions. 
While the Gaussian-pDP yields distributions with lower Wasserstein Distances on most datasets when privacy preservation levels are low, its accuracy decreases drastically with even slight increases in privacy levels. 
Similarly, the Laplace-DP and Flipped Huber demonstrate competitive precision in data distribution at lower privacy levels but undergo steep declines in accuracy as privacy levels rise. 
This is due to the noise-addition-based data perturbation methods employed by the Gaussian-pDP, Laplace-DP, and Flipped Huber. 
When privacy levels are low, the noise added is close to zero, resulting in higher distribution accuracy. 
However, as the degree of privacy preservation increases, the variance of the noise expands, drastically degrading the precision of the data distributions. 
In contrast, our RVNS maintains the highest accuracy in data distributions at relatively high privacy levels while remaining competitive at lower privacy levels. 
Conversely, the EMS consistently exhibits lower accuracy in data distribution estimation across all privacy levels.

\begin{figure*}[!t]
	\centering
	\subfloat[$\chi^2(2)$]{\includegraphics[width=0.23\linewidth]{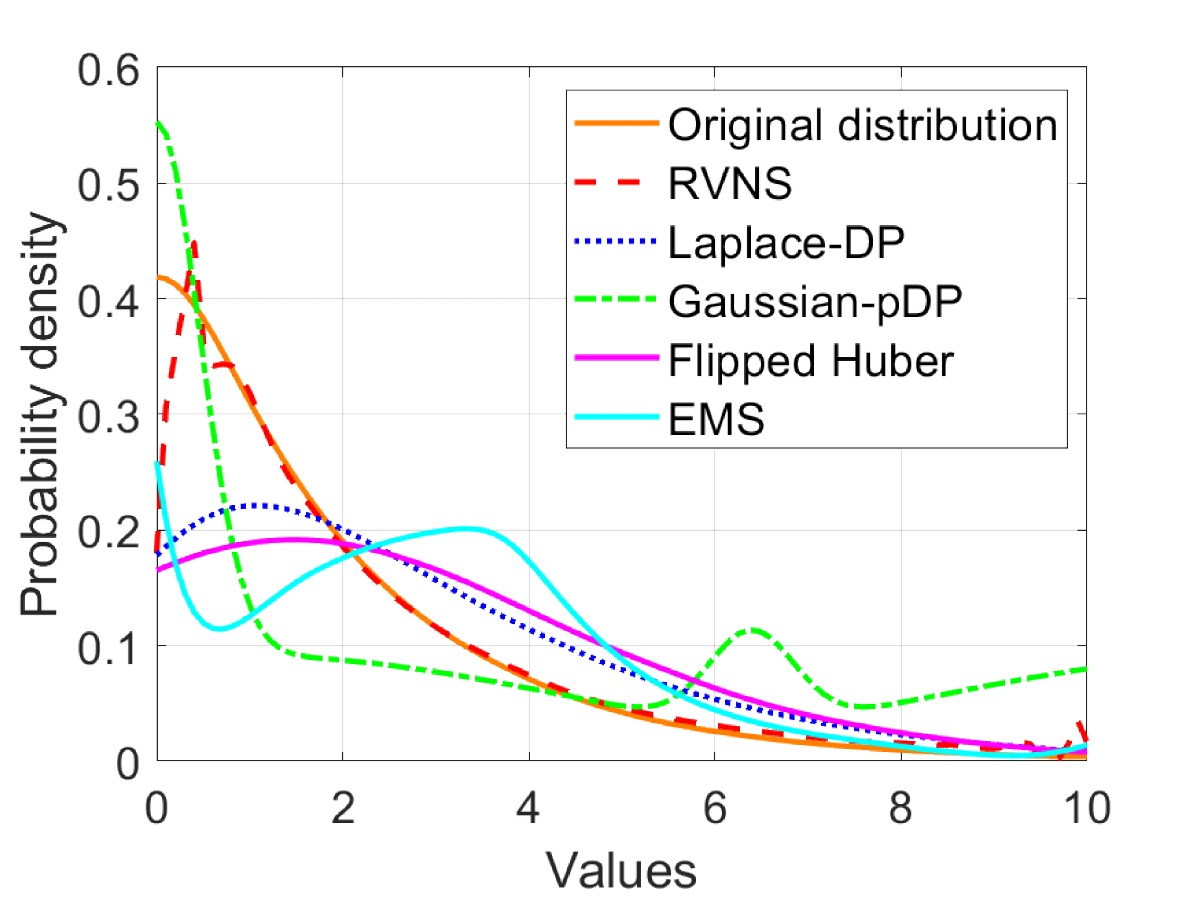}}
	\hfil
	\subfloat[$\chi^2(3)$]{\includegraphics[width=0.23\linewidth]{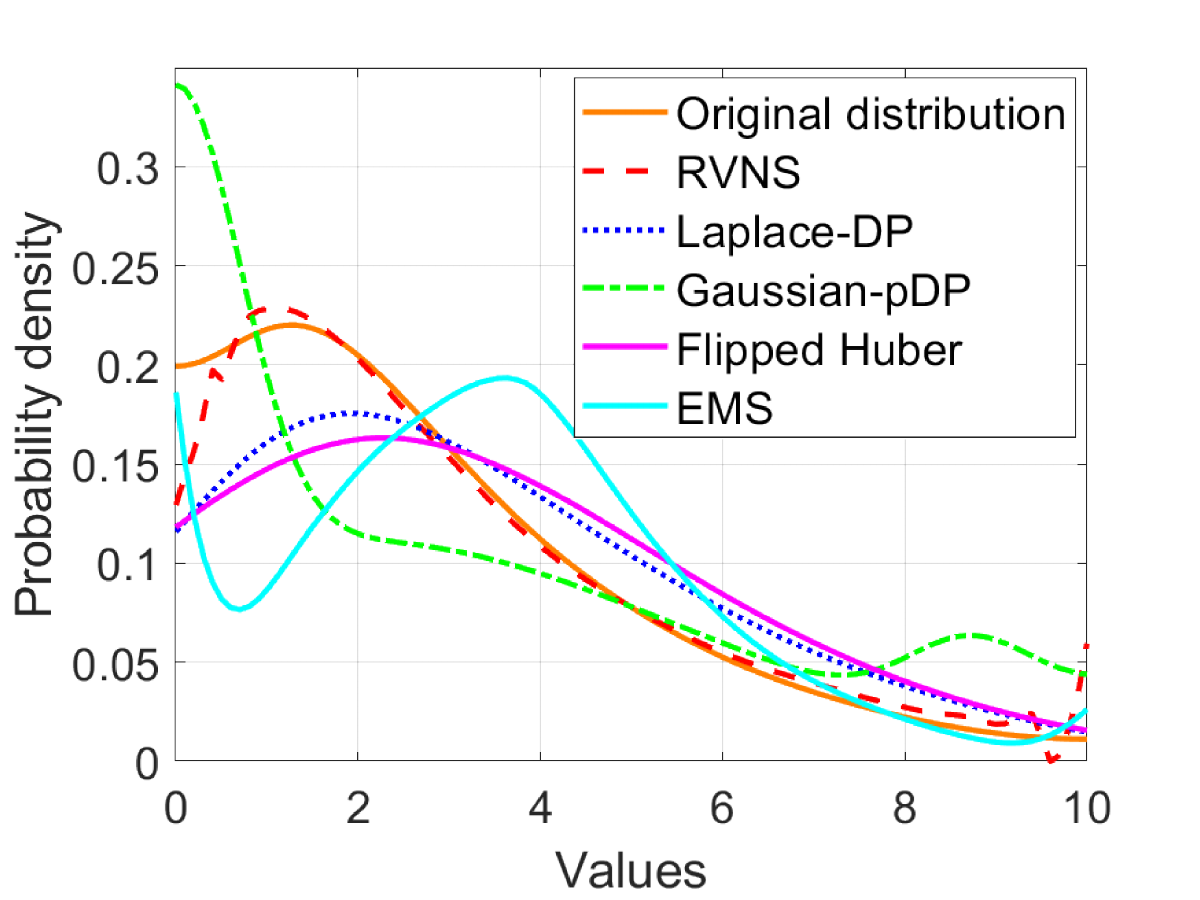}}
	\hfil
	\subfloat[$\chi^2(5)$]{\includegraphics[width=0.23\linewidth]{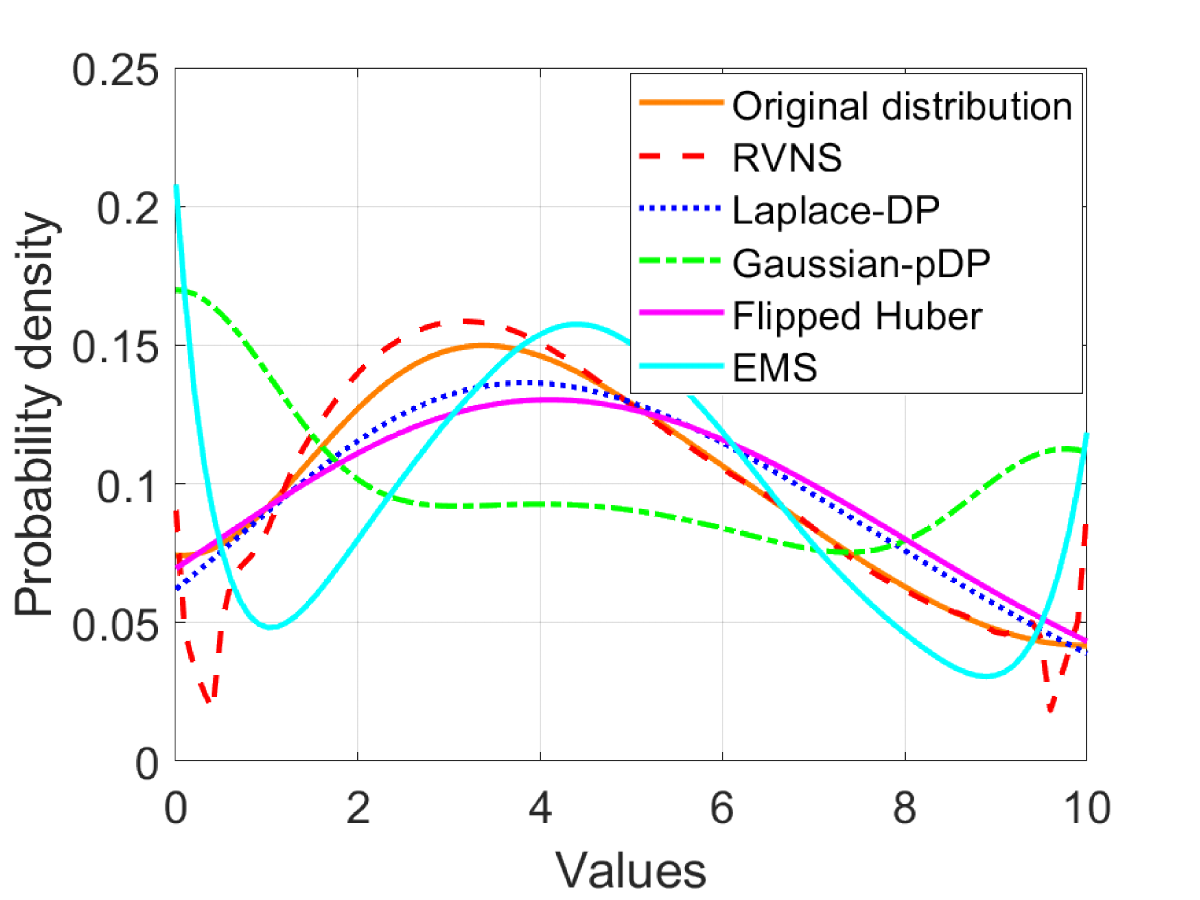}}
	\hfil
	\subfloat[$\chi^2(8)$]{\includegraphics[width=0.23\linewidth]{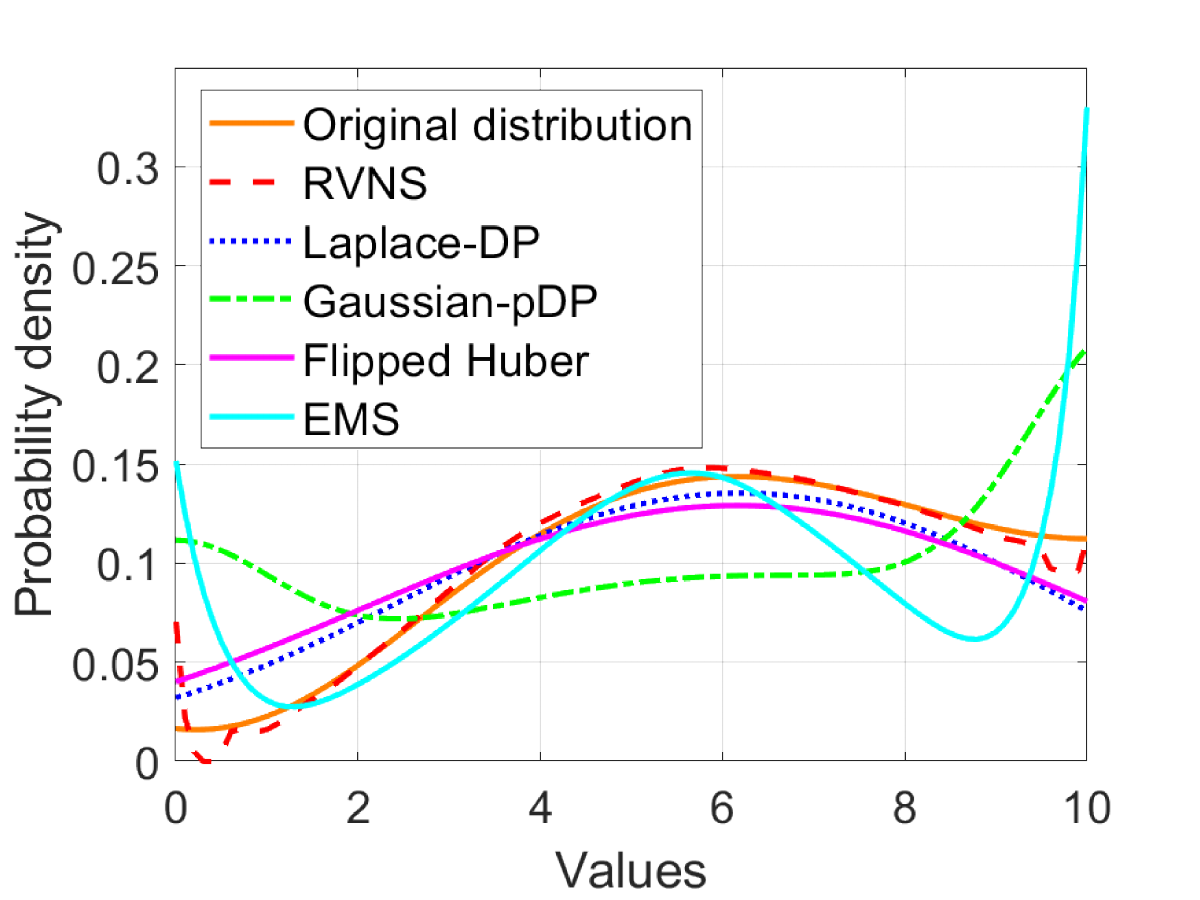}}
	\hfil
	\subfloat[$\chi^2(10)$]{\includegraphics[width=0.23\linewidth]{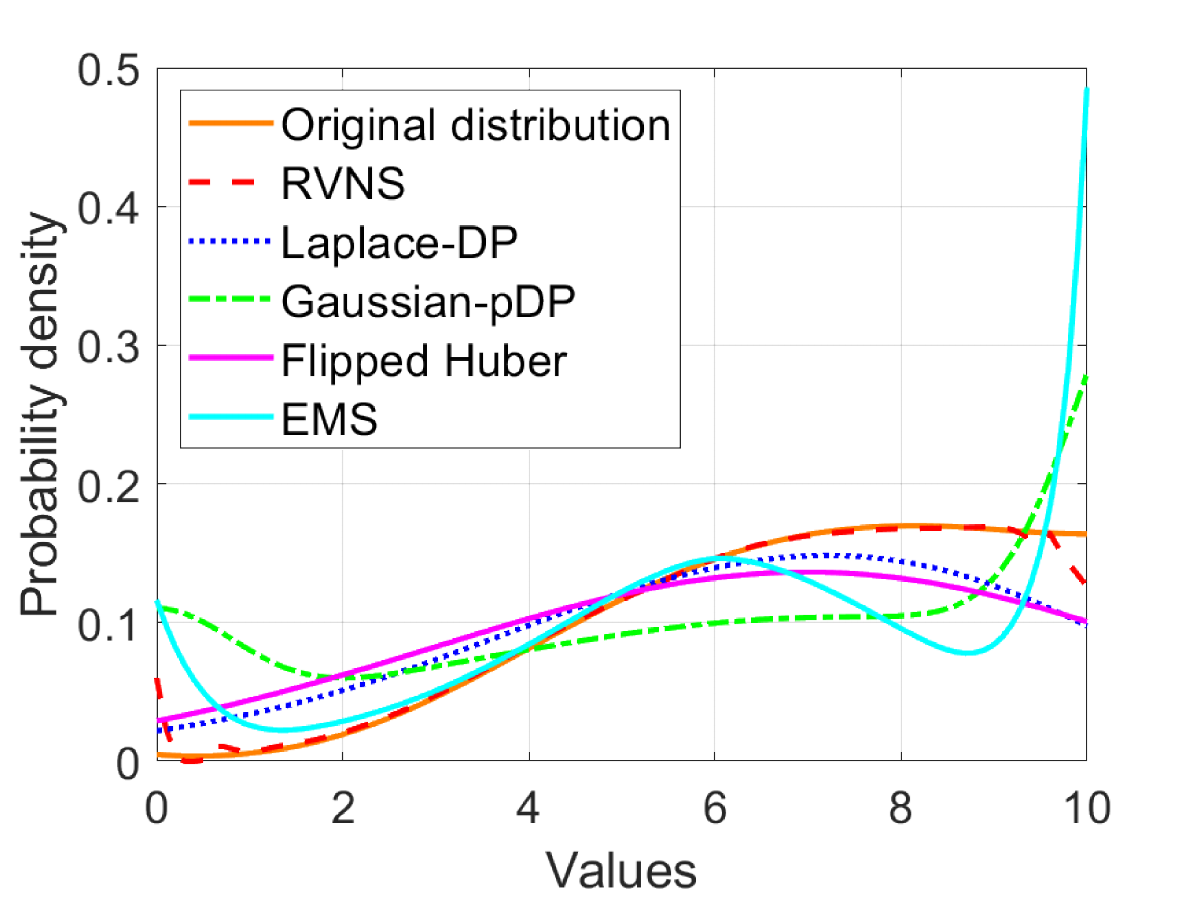}}
	\hfil
	\subfloat[BMI]{\includegraphics[width=0.23\linewidth]{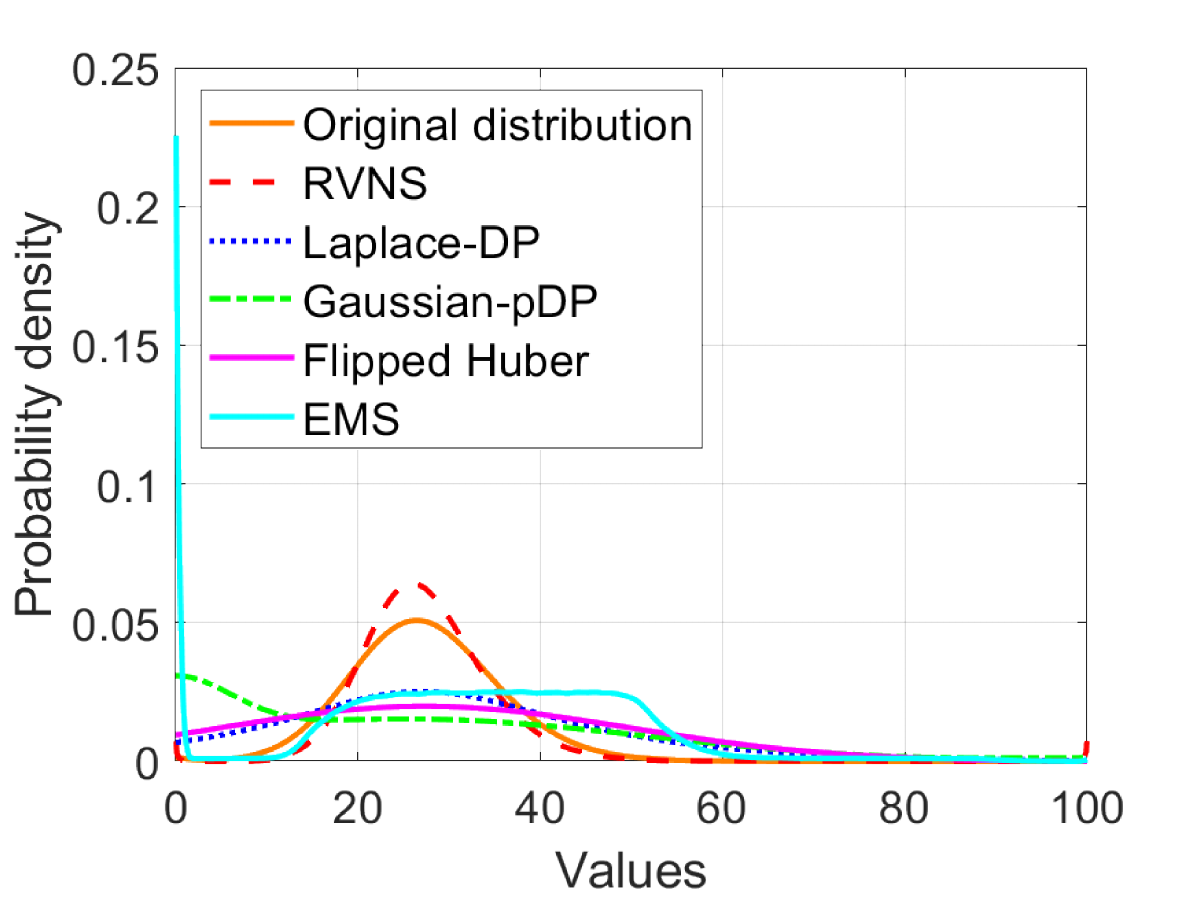}}
	\hfil
	\subfloat[Bagrut Grade]{\includegraphics[width=0.23\linewidth]{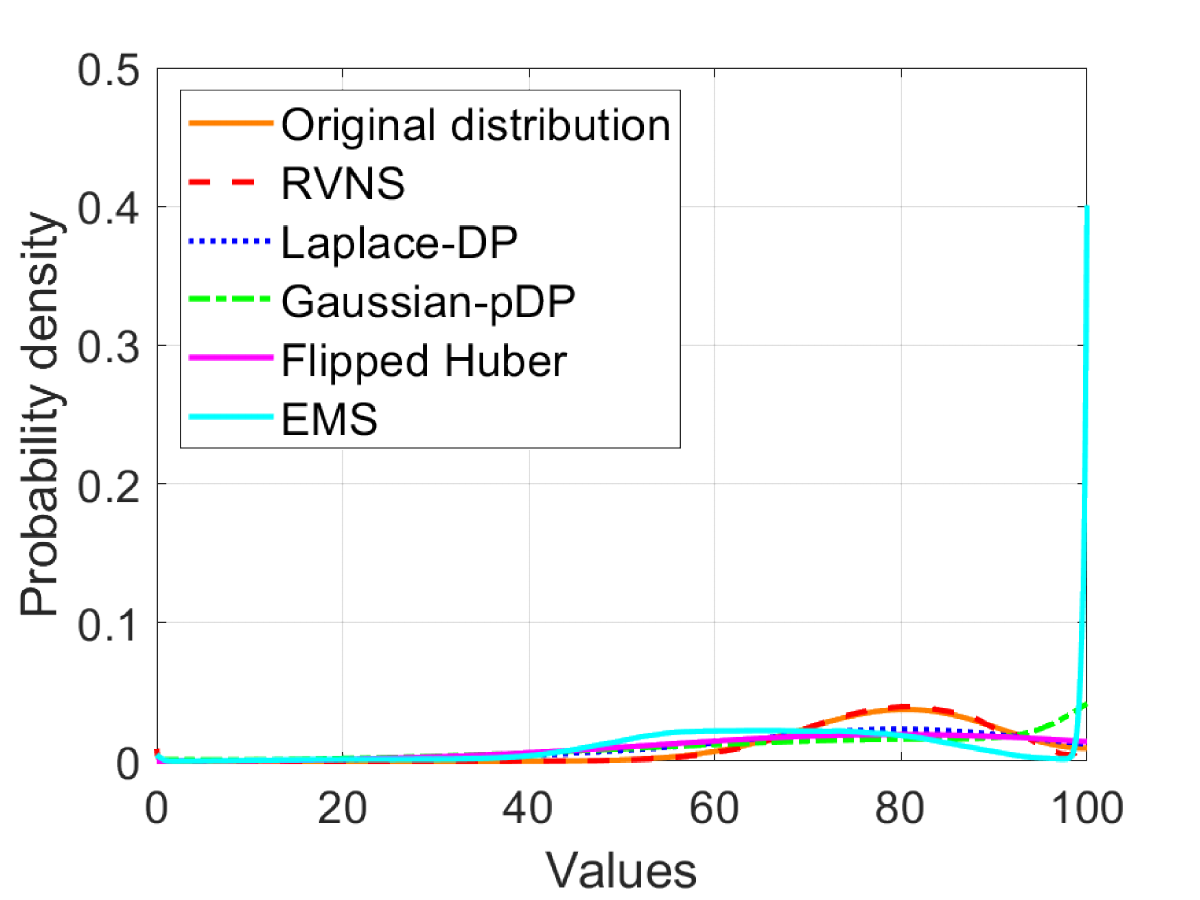}}
	\hfil
	\subfloat[IMDB Rating]{\includegraphics[width=0.23\linewidth]{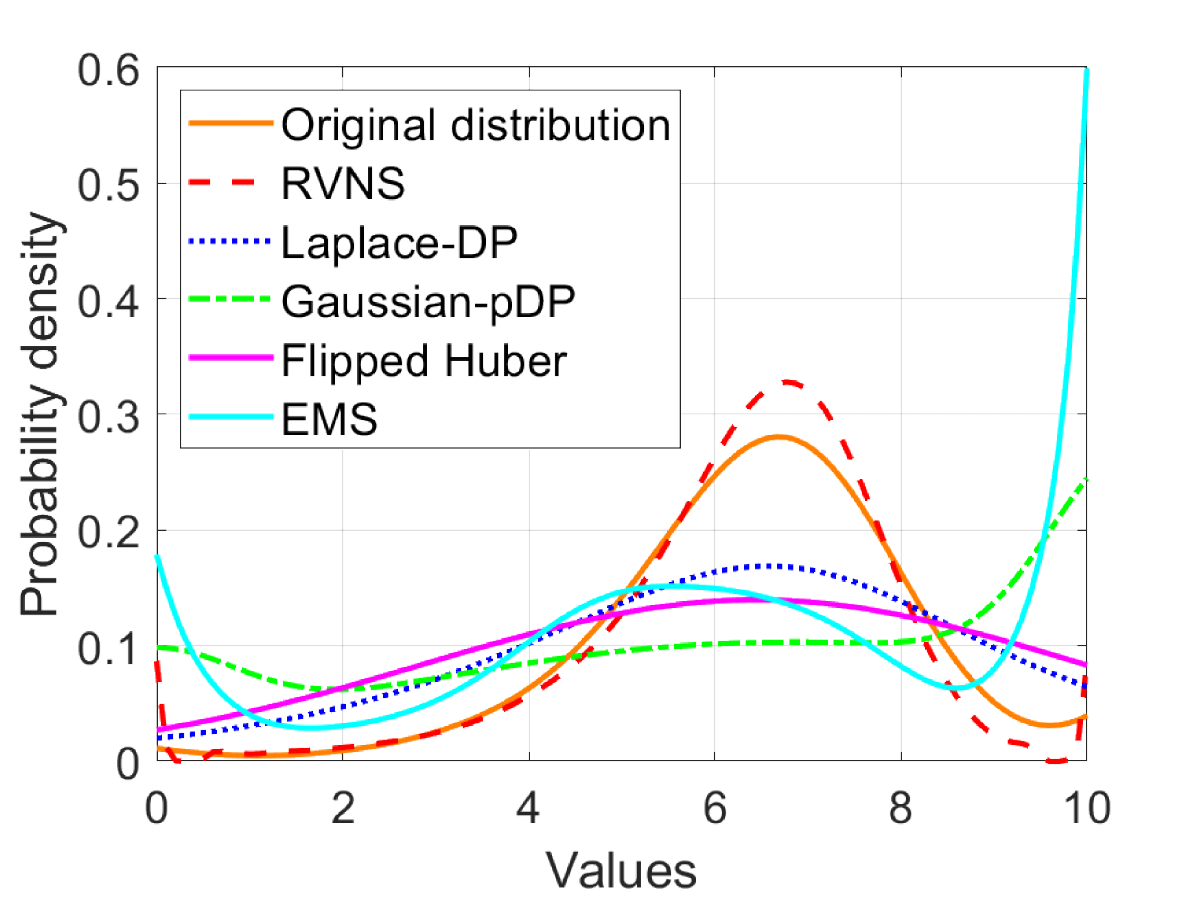}}	
	\caption{The probability density of data perturbed by the RVNS and four compared algorithms on five synthetic datasets and three real-world datasets.}
	\label{fig:KDE}
\end{figure*}

Secondly, in scenarios where data utility remains constant, the RVNS can offer superior levels of privacy preservation. 
Fig. \ref{fig:WE} demonstrates that the accuracy of data distributions produced by the four existing comparison algorithms decreases sharply as privacy preservation levels increase. 
Notably, although the Gaussian-pDP achieves the highest accuracy in data distribution at lower privacy levels, its accuracy drops drastically as privacy levels rise, eventually becoming the worst among all algorithms. 
In contrast, the accuracy of data distributions obtained through the RVNS remains largely unchanged with increasing privacy levels. 
This is because the RVNS does not require discretization of the original data and fully considers the similarity in distribution between the reconstructed data and the original data during the reconstruction process. 
Therefore, under high levels of privacy preservation, the RVNS is capable of obtaining high-utility data distributions. 
Thus, when users demand stringent privacy preservation, the RVNS emerges as the preferred method for obtaining more accurate data distributions.

To more intuitively show the accuracy of data distribution obtained by different methods, we selected experimental results where the privacy level was near the midpoint in the trade-off across different datasets, as depicted in Fig. \ref{fig:WE}. 
Then, we utilized the KDE method to ascertain the probability density of the data distributions obtained by these methods and presented them in Fig. \ref{fig:KDE}.
The privacy and utility corresponding to the selected experimental results are provided in the supplementary material.

Analysis of Fig. \ref{fig:KDE} reveals that the probability density estimated by the RVNS provides a relatively precise reflection of the changes in the original data's probability density. 
Specifically, on synthetic datasets, while the Gaussian-pDP, Laplace-DP, and Flipped Huber roughly capture the location of the original probability density's maximum value, the disparities between their maxima and the original maximum are significantly higher than those observed with the RVNS. 
Furthermore, the location of the maximum identified by the RVNS is more precise. 
On real-world datasets, the Gaussian-pDP, Laplace-DP, and Flipped Huber struggle to locate the original probability density's maximum value. 
Conversely, the RVNS consistently and accurately identifies the position of the original probability density's maximum value, as well as attains the highest probability value. 
Regarding the EMS, it fails to locate the maximum value's position in both synthetic and real-world data.

\begin{figure*}[!t]
	\centering
	\subfloat[$Mean$]{\includegraphics[width=0.32\linewidth]{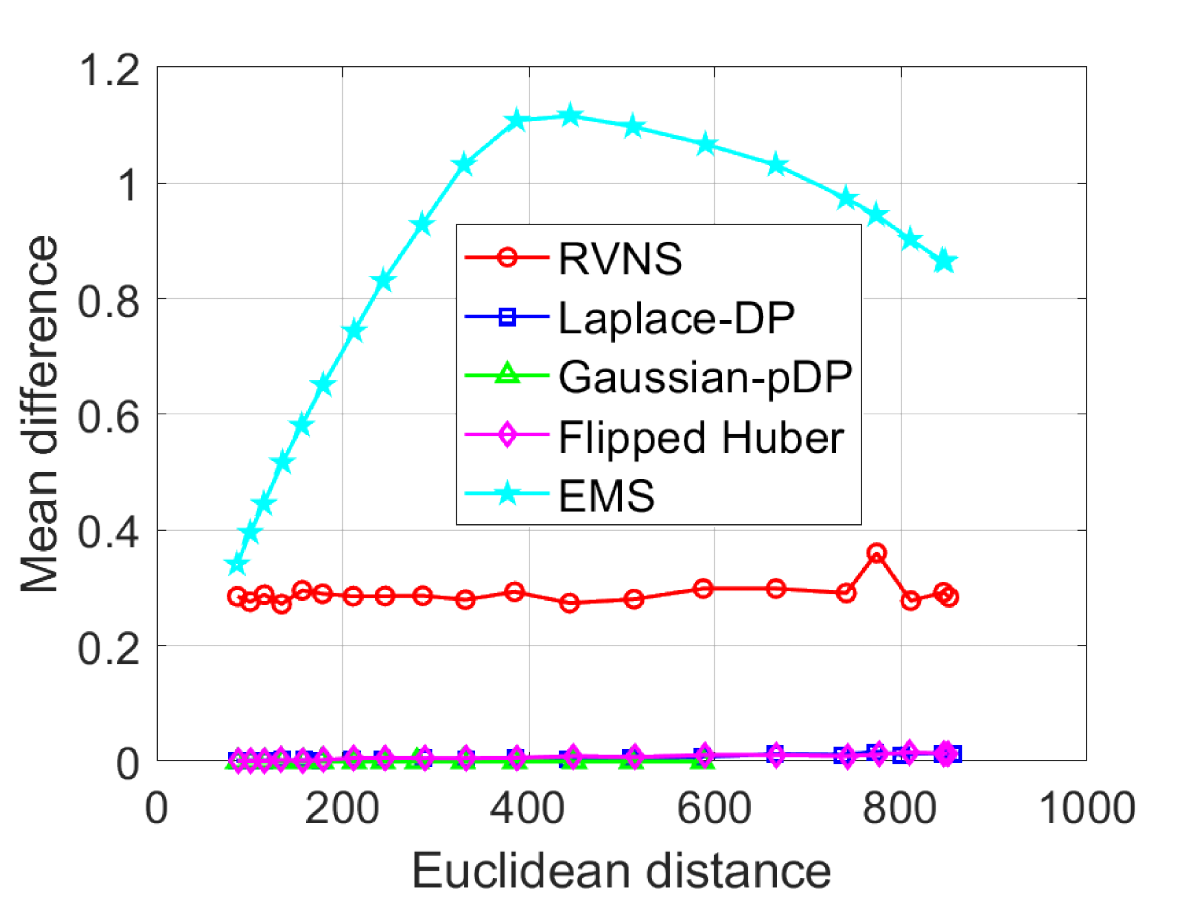}}
	\hfil
	\subfloat[$Standard\ Deviation$]{\includegraphics[width=0.32\linewidth]{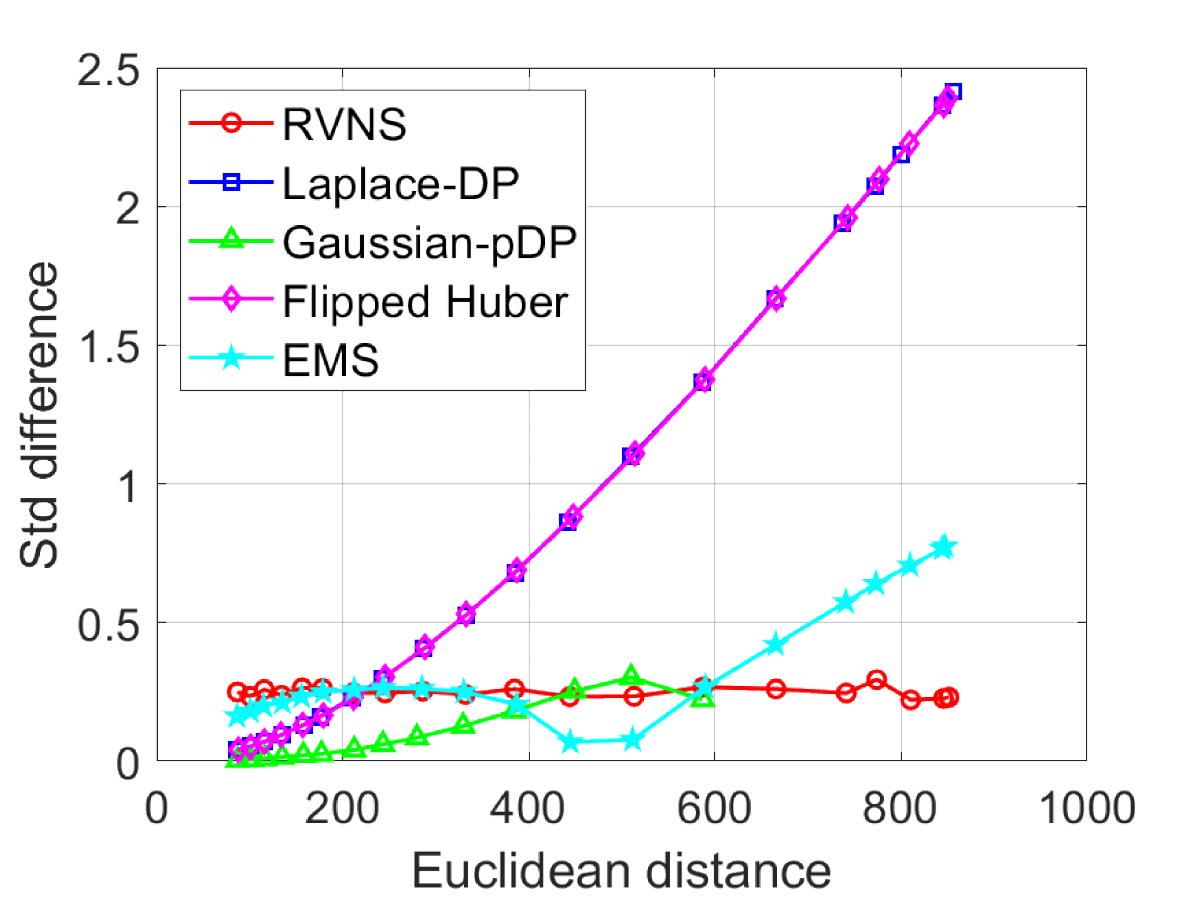}}
	\hfil
	\subfloat[$Mode$]{\includegraphics[width=0.32\linewidth]{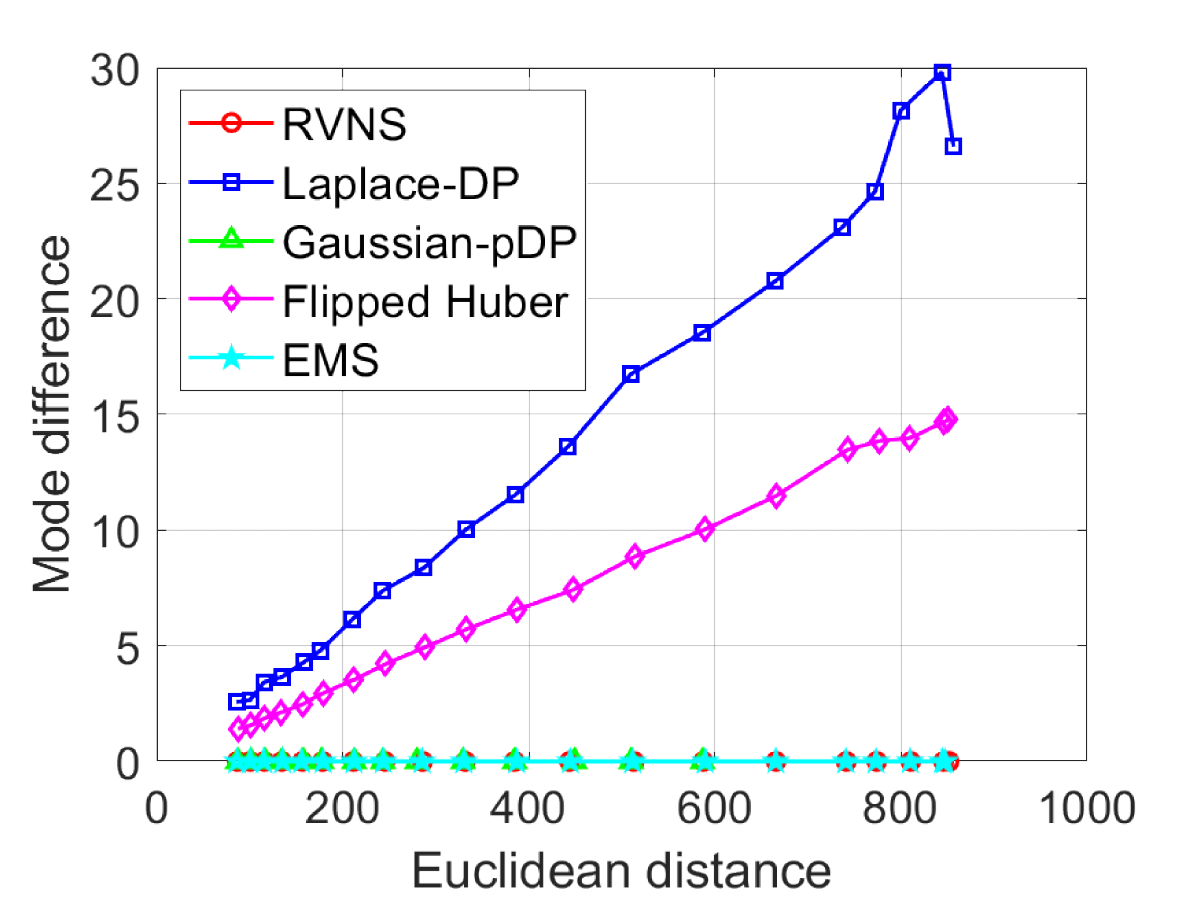}}
	\hfil
	\subfloat[$Median$]{\includegraphics[width=0.32\linewidth]{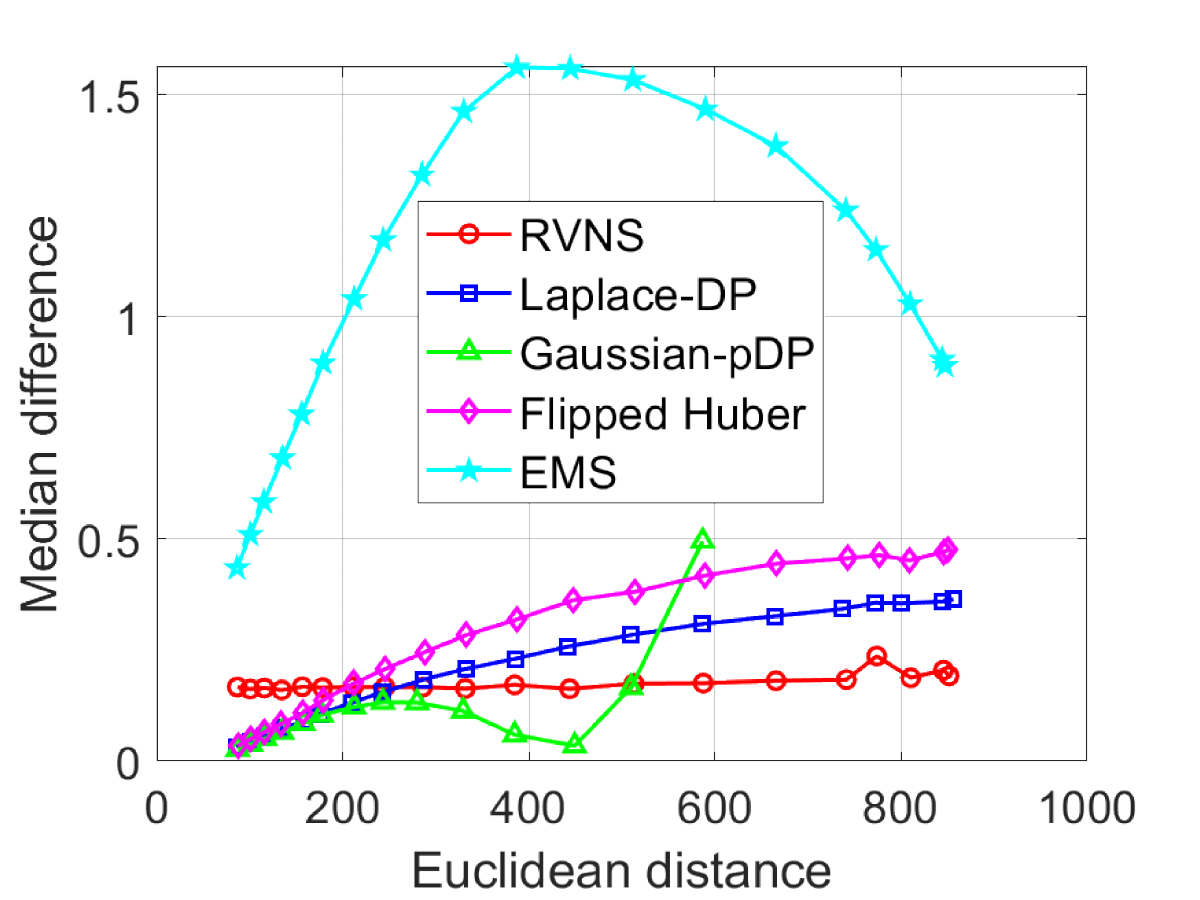}}
	\hfil
	\subfloat[$Skewness$]{\includegraphics[width=0.32\linewidth]{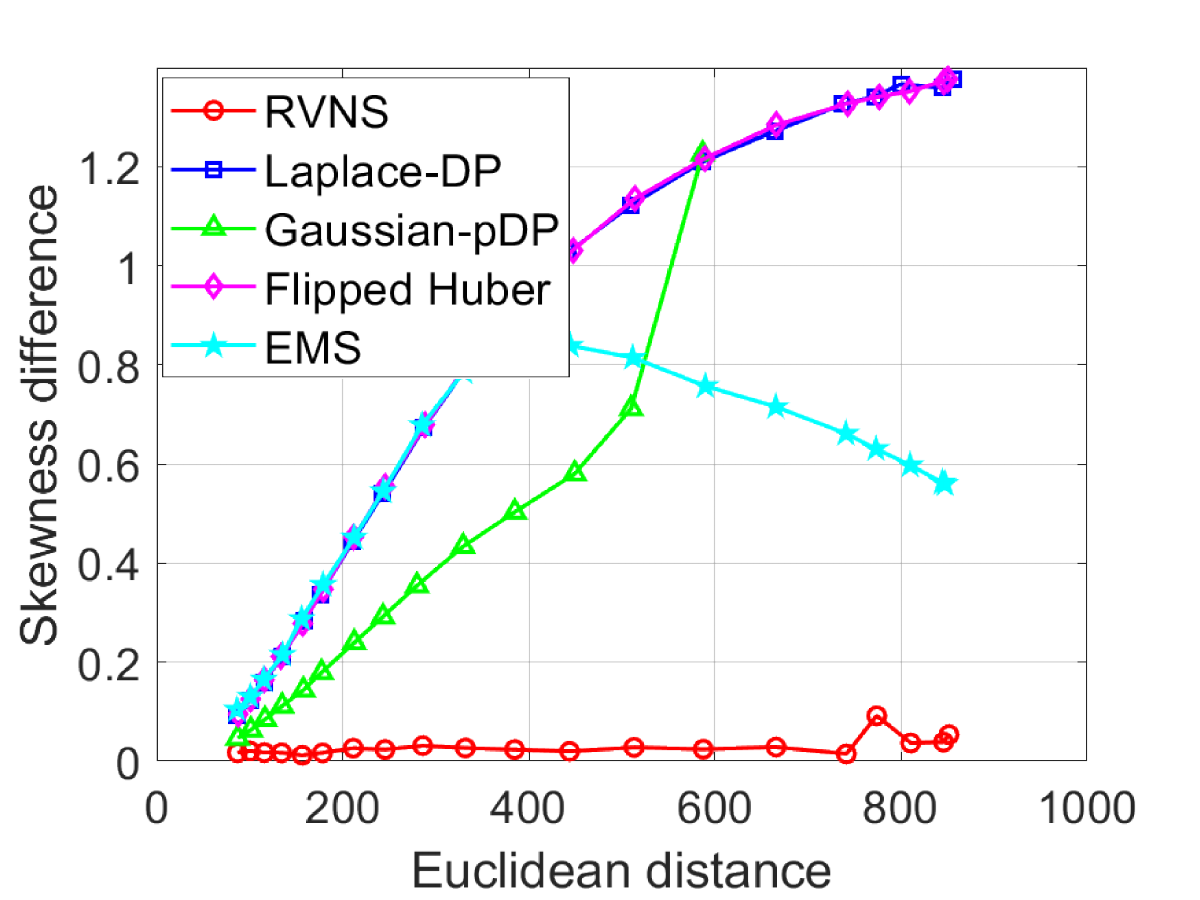}}
	\hfil
	\subfloat[$Kurtosis$]{\includegraphics[width=0.32\linewidth]{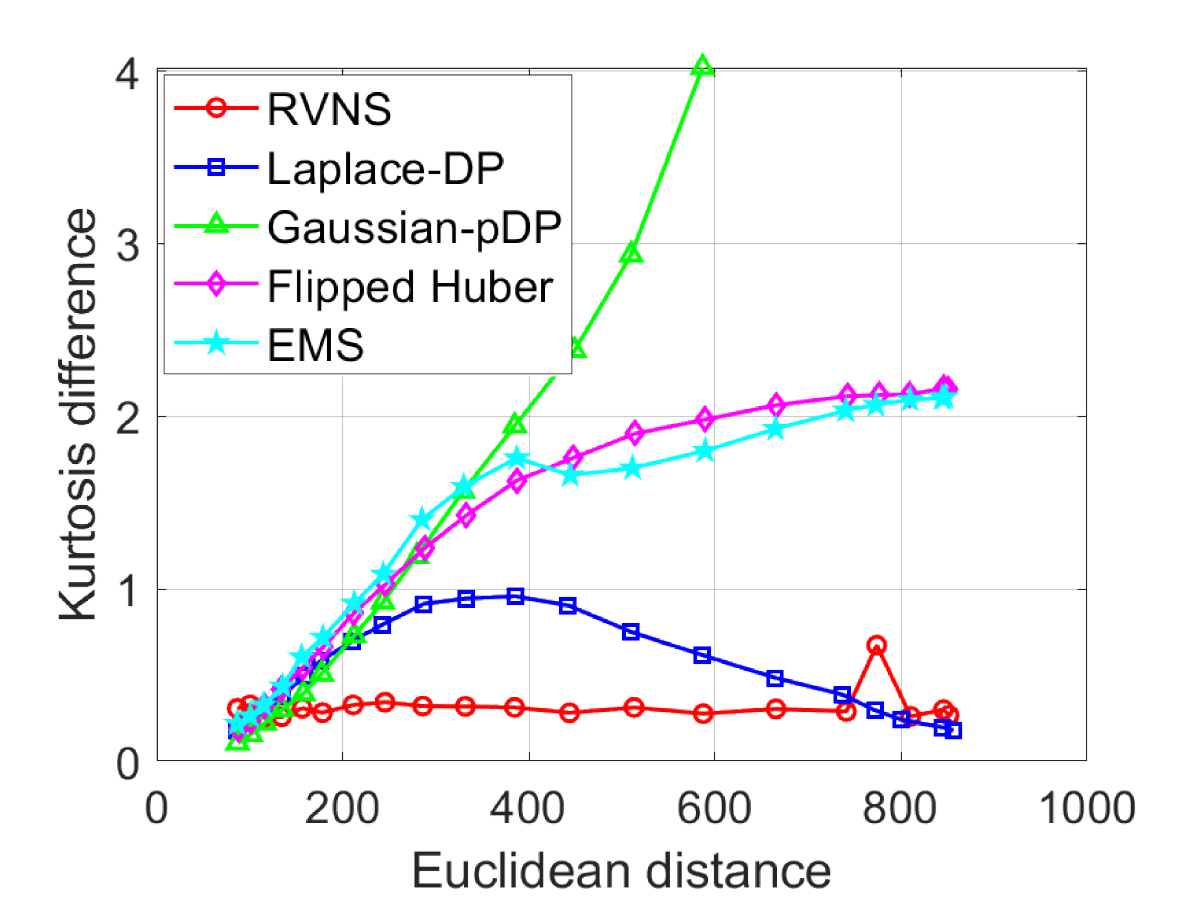}}
	
	\caption{The six statistical indicators of data perturbed by the RVNS and four compared algorithms on synthetic dataset $\chi^2(2)$.}
	\label{fig:six-chi2}
\end{figure*}

It is noteworthy that, in Fig. \ref{fig:KDE}, the probability density of both the RVNS and EMS increases sharply near the boundaries of the value range. 
This phenomenon stems from the fact that both RVNS and EMS impose constraints on the generation range of perturbed data, and they utilize KDE to obtain the probability density distribution of the perturbed data. 
Due to the boundary effect of the KDE estimation method \cite{jones1993simple}, the perturbed data at the boundaries have lower probability densities, leading to a sudden increase in the reconstructed probability density at the boundaries. 
In contrast, the Gaussian-pDP, Laplace-DP, and Flipped Huber do not restrict the range of values for the perturbed data. 
Although their results are also subject to KDE-based probability density estimation, their perturbed data's range significantly exceeds the range used for estimation, thereby avoiding the sharp increase in probability density near the value range boundaries.
Although the results obtained by the RVNS have boundary effects, the probability density obtained by it is still the most accurate.

\subsubsection{Six Statistical Indicators}
\begin{figure*}[!t]
	\centering
	\subfloat[$Mean$]{\includegraphics[width=0.32\linewidth]{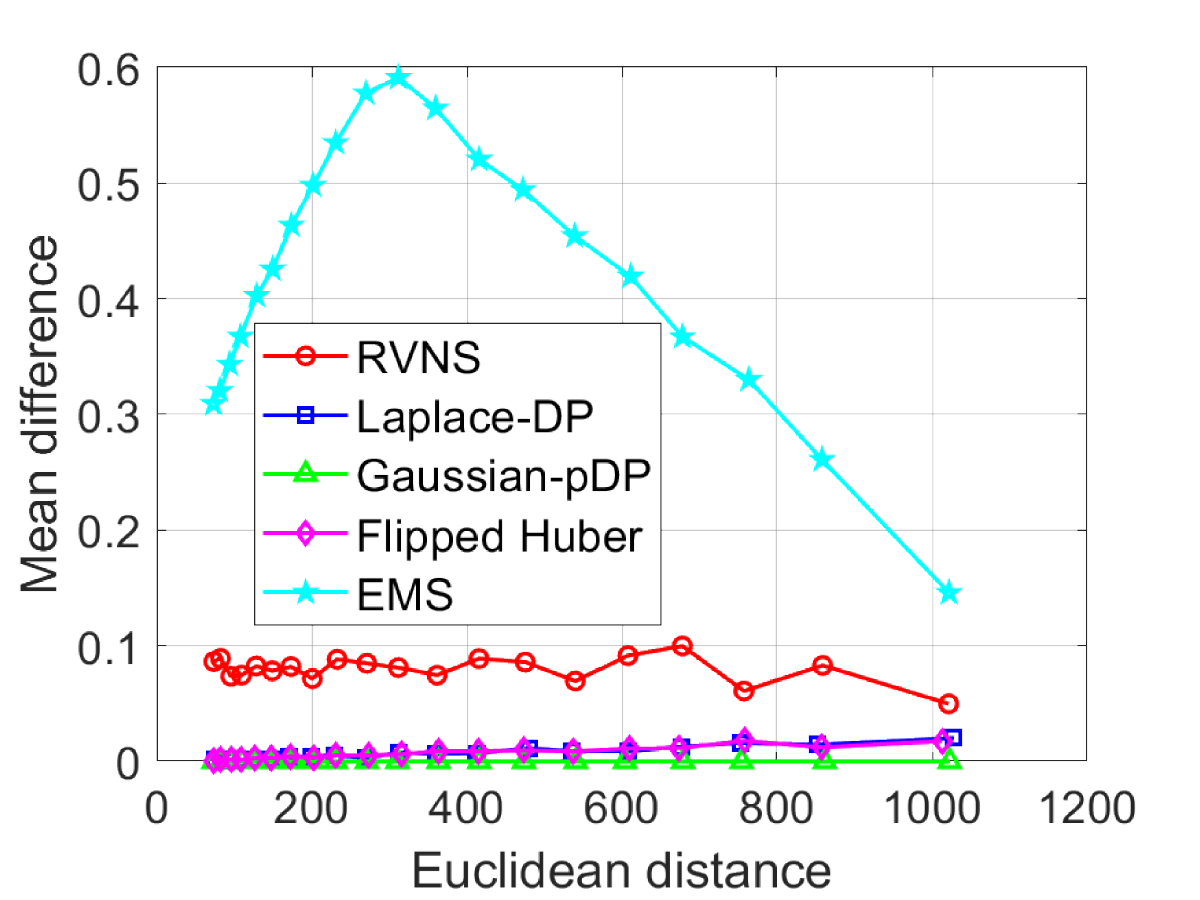}}
	\hfil
	\subfloat[$Standard\ Deviation$]{\includegraphics[width=0.32\linewidth]{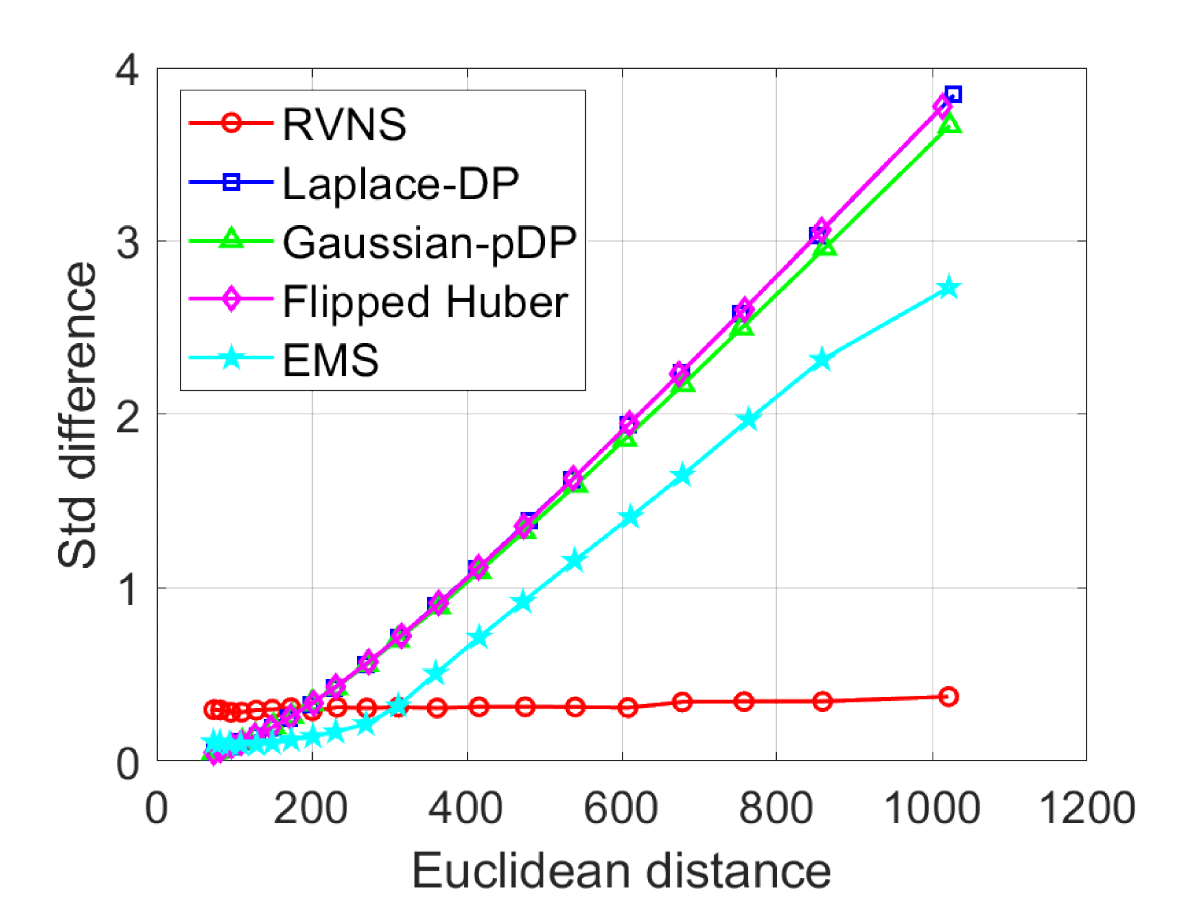}}
	\hfil
	\subfloat[$Mode$]{\includegraphics[width=0.32\linewidth]{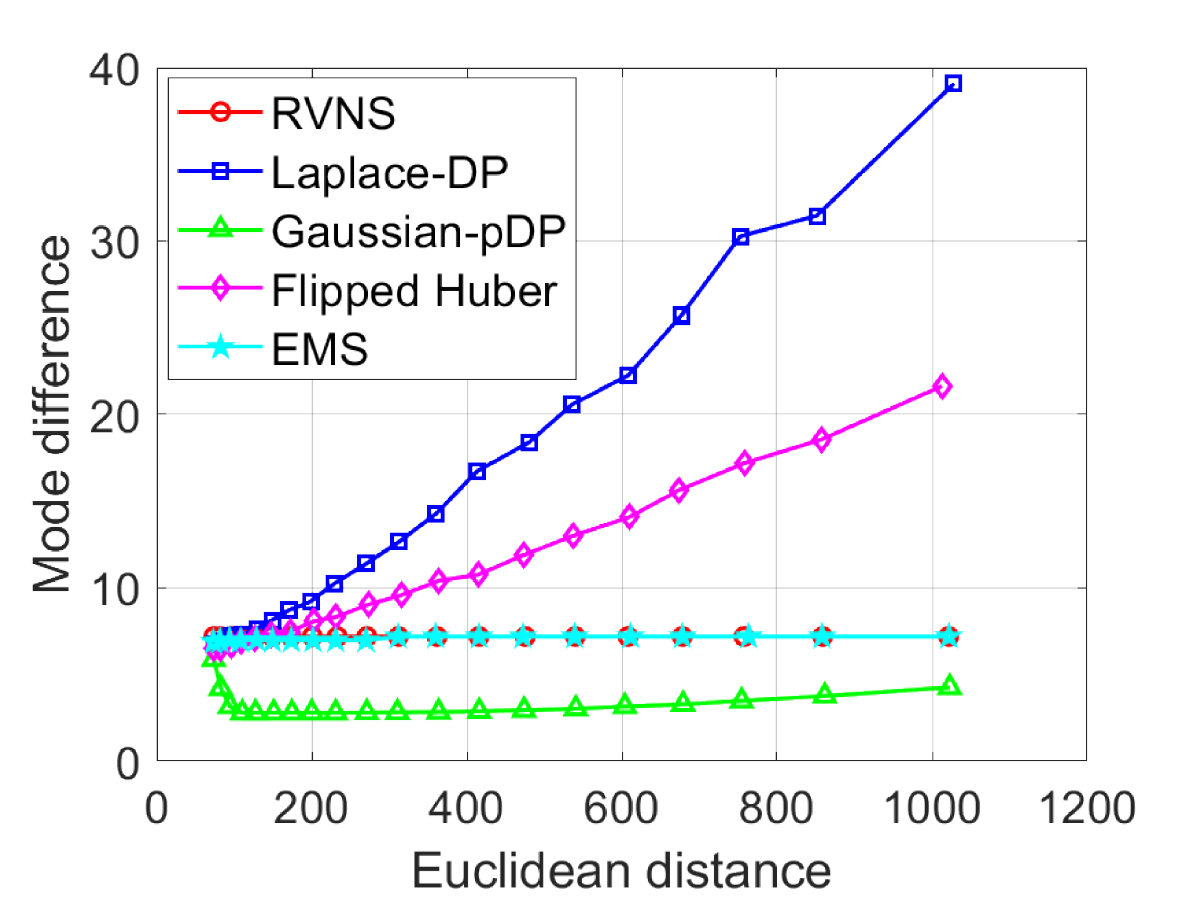}}
	\hfil
	\subfloat[$Median$]{\includegraphics[width=0.32\linewidth]{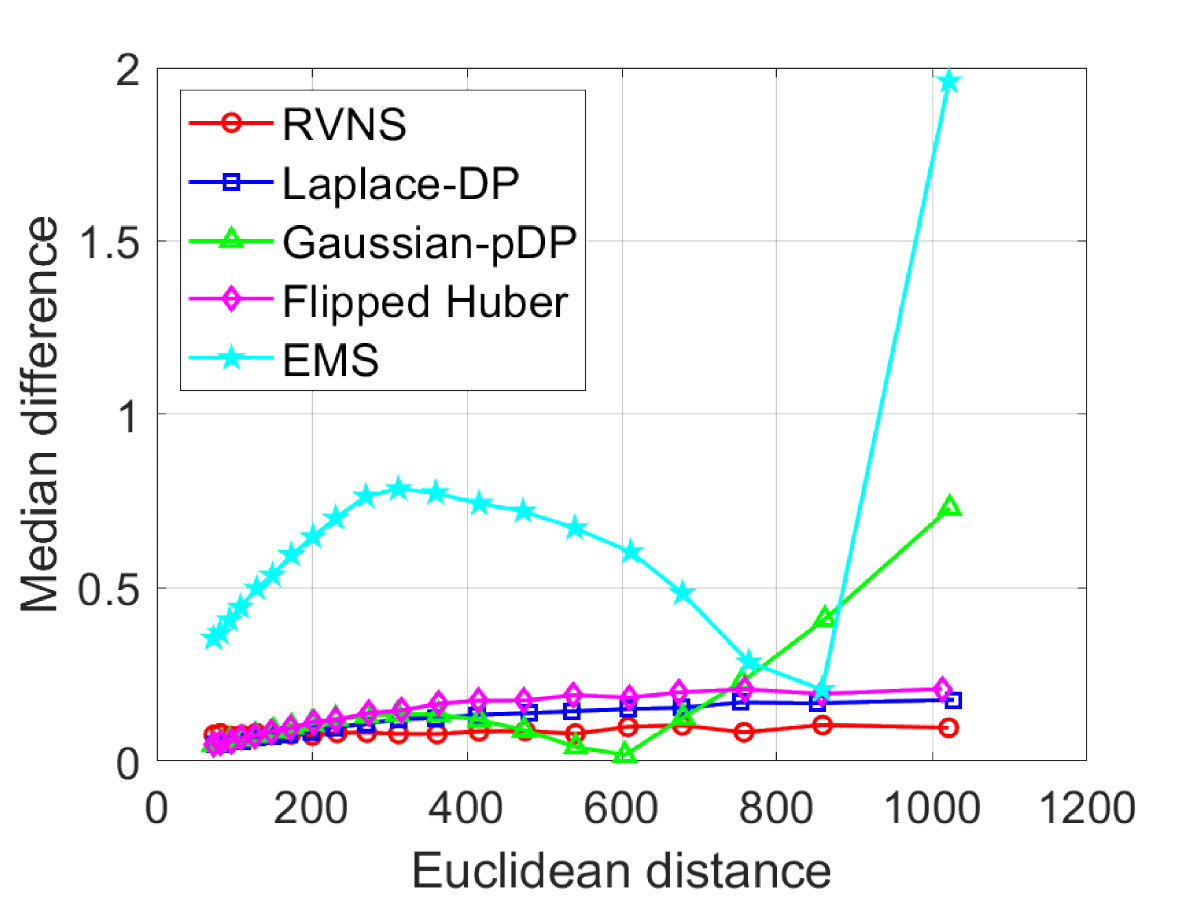}}
	\hfil
	\subfloat[$Skewness$]{\includegraphics[width=0.32\linewidth]{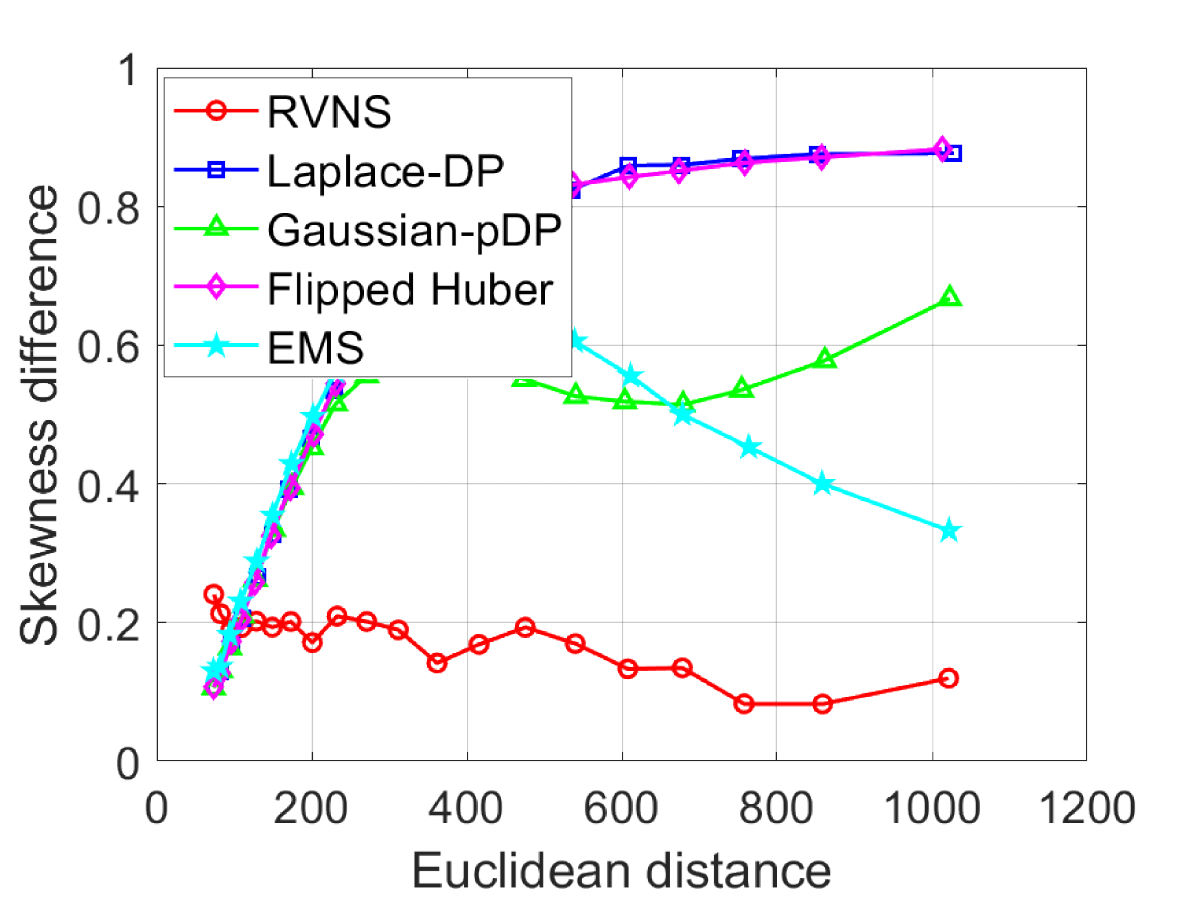}}
	\hfil
	\subfloat[$Kurtosis$]{\includegraphics[width=0.32\linewidth]{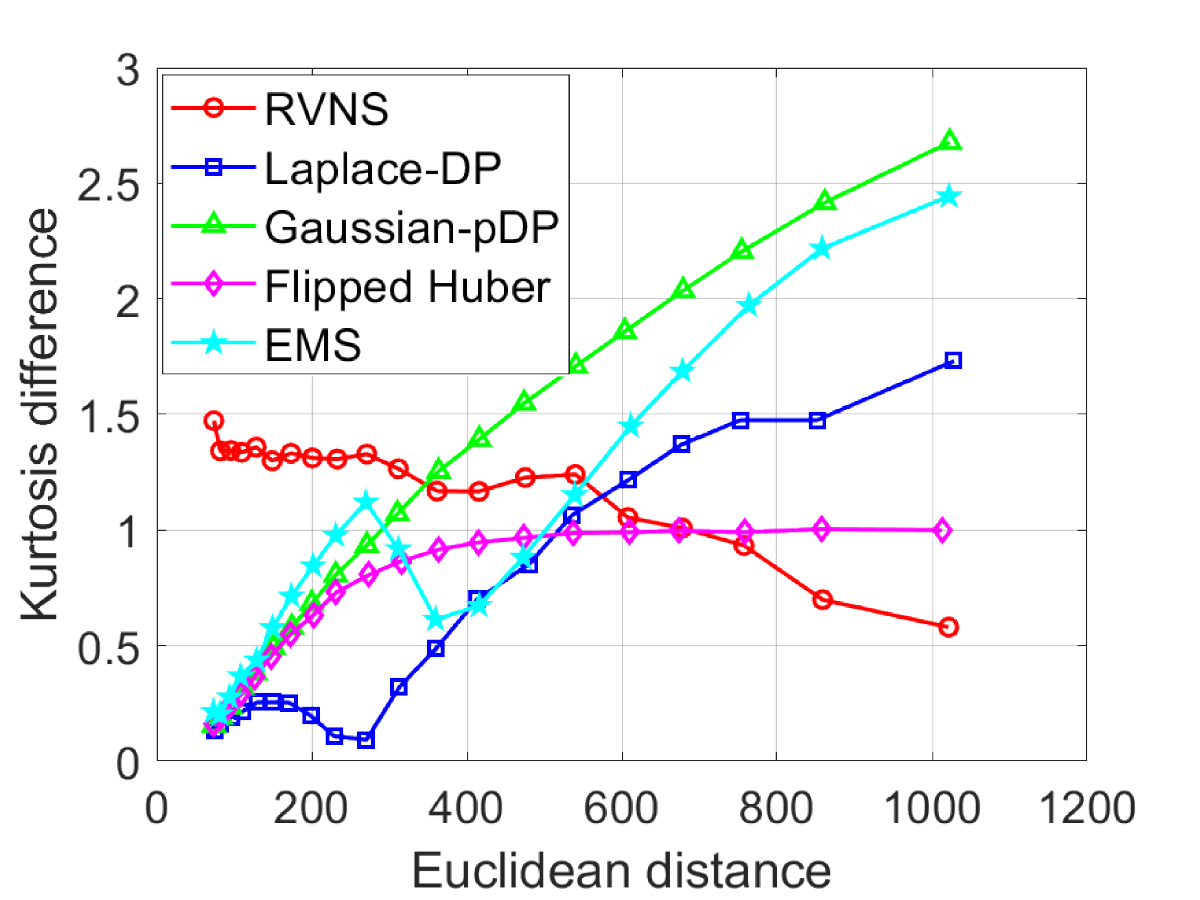}}
	
	\caption{The six statistical indicators of data perturbed by the RVNS and four compared algorithms on real-world dataset IMDB Rating.}
	\label{fig:six-movies}
\end{figure*}

The experimental results are presented in Figs. \ref{fig:six-chi2}-\ref{fig:six-movies} offer comprehensive insights into the efficacy of the RVNS in supporting diversified statistical analysis across various datasets ($\chi^2(2)$ and IMDB Rating). 
The analysis focuses on six indicators, including $Mean$, $Standard\ Deviation$, $Mode$, $Median$, $Skewness$, and $Kurtosis$, each of which is examined about varying levels of privacy.
Due to the page limitation, the efficacy of the RVNS in supporting diversified statistical analysis across datasets $\chi^2(5)$ and $\chi^2(10)$ is provided in the supplementary material.

In Fig. \ref{fig:six-chi2}, the data derived from the RVNS demonstrates an overall superior accuracy in statistical analysis. 
The Gaussian-pDP, while achieving high precision in Mean and $Standard\ Deviation$, exhibits significant deviations from the original results in the remaining four indicators. 
The Laplace-DP and Flipped Huber exhibit satisfactory performance solely in Mean, with their $Standard\ Deviation$ results maintaining high accuracy only at lower privacy levels, but declining sharply as privacy increases. 
The EMS, on the other hand, achieves accurate $Standard\ Deviation$ statistics only within a privacy range of 400-600. 
By contrast, the RVNS consistently maintains high precision across all six indicators, delivering accurate statistical results in most scenarios.

Finally, Fig. \ref{fig:six-movies} demonstrates that the RVNS produces the most accurate statistically analyzed perturbed data even on real-world datasets.
The Gaussian-pDP surpasses the RVNS in the Mean and $Mode$ but falls behind in the other four indicators. 
The Laplace-DP and Flipped Huber are competitive only in the Mean and $Kurtosis$, with inferior performance in the other indicators. 
As for the EMS, it does not exceed the RVNS in any of the six indicators.

Collectively, the experimental results presented in Figs. \ref{fig:six-chi2}-\ref{fig:six-movies} affirm that data obtained by the RVNS better supports diversified statistical analysis than the other compared algorithms, exhibiting consistent and superior accuracy across various datasets and privacy levels.

\section{Conclusion and future work}
\label{sec:conclu}
In this study, we have presented a real-value negative survey model, called RVNS, specifically designed to capture the comprehensive distribution of users' real-valued data while ensuring the confidentiality of their sensitive personal information. 
Our analysis, supported by both theoretical insights and experimental evidence, underscores the efficacy of the proposed RVNS model in offering robust protection for individuals' data privacy while simultaneously achieving a precise depiction of the data distribution. 
This dual capability furnishes vital data support for a wide array of subsequent statistical analyses.

Looking toward the future, we anticipate extending the applicability of RVNS to multidimensional real-valued data, thereby expanding its potential in various data privacy and statistical analysis contexts. 
Furthermore, the exploration of methods to offer differentiated levels of privacy protection tailored to individual users emerges as a promising avenue for future research. 
Such advancements will facilitate even more granular and customized privacy-preserving data analysis, ultimately enhancing both the security and utility of sensitive real-valued data across diverse applications.

%\section{Conclusion}
%The conclusion goes here.
%
%
%\section*{Acknowledgments}
%This should be a simple paragraph before the References to thank those individuals and institutions who have supported your work on this article.

\bibliographystyle{IEEEtran}
\bibliography{mybibfile}

\end{document}